\documentclass{aa}
\usepackage{savesym}
\savesymbol{bibfont}
\usepackage{amsmath}
\usepackage{lscape}
\usepackage{rotating}
\usepackage{natbib}
\usepackage{graphicx}
\usepackage{txfonts}
\setcitestyle{aysep={}}
\restoresymbol{NAT}{bibfont}

\begin{document}

\title{The X-ray light curve of gamma-ray bursts: clues to the central engine}

\author{
M.G. Bernardini\inst{1,2}
\and
R. Margutti\inst{1,3}
\and
J. Mao\inst{4,5}
\and
E. Zaninoni\inst{1,6}
\and
G. Chincarini\inst{1,7}
}

\institute{
INAF - Osservatorio Astronomico di Brera, via Bianchi 46, I-23807 Merate (LC), Italy 
\and
ICRANet, p.le della Repubblica 10, I-65100 Pescara, Italy
\and
Harvard-Smithsonian Center for Astrophysics, 60 Garden Street, Cambridge, MA02138, USA
\and
Yunnan Observatory \& Key Laboratory for the Structure and Evolution of Celestial Objects, Chinese Academy of Sciences, Kunming, Yunnan Province, 650011, China 
\and
International Center for Astrophysics,  Korean Astronomy and Space Science Institute, 776, Daedeokdae-ro, Yuseong-gu, Daejeon, Republic of Korea
\and
University of Padova, Astronomy Dept., v. dell'Osservatorio 3, I-35122 Padova, Italy
\and
University of Milano  Bicocca, Physics Dept., P.zza della Scienza 3, Milano 20126, Italy
}

\titlerunning{X-ray emission of gamma-ray bursts}

\authorrunning{Bernardini et al.}

\date{}

\abstract
{}
{We present the analysis of a large sample of gamma-ray burst (GRB) X-ray light curves in the rest frame to characterise their intrinsic properties in the context of different theoretical scenarios.}
{We determine the morphology, time scales, and energetics of $64$ long GRBs observed by \emph{Swift}/XRT \emph{without} flaring activity. We furthermore provide a one-to-one comparison to the properties of GRBs \emph{with} X-ray flares.}
{We find that the steep decay morphology and its connection with X-ray flares favour a scenario in which a central engine origin. We show that this scenario can also account for the shallow decay phase, provided that the GRB progenitor star has a self-similar structure with a constant envelope-to-core mass ratio $\sim 0.02-0.03$. However, difficulties arise for very long duration ($t_p\gtrsim10^4$ s) shallow phases. Alternatively, a spinning-down magnetar whose emitted power refreshes the forward shock can quantitatively account for the shallow decay properties. In particular we demonstrate that this model can account for the plateau luminosity vs. end time anticorrelation.}
{}


\maketitle

\section{Introduction}

Since the launch of the \emph{Swift} satellite in 2004 \citep{2004ApJ...611.1005G}, the evolution of the gamma-ray burst (GRB) X-ray light curves has appeared to be more complex than previously thought. The different shapes of the light curves \citep{2006ApJ...642..389N} and the existence of X-ray flares superimposed on the smooth continuum \citep{2007ApJ...671.1903C,chinca10} up to very late times \citep{2008A&A...487..533C,2011A&A...526A..27B} have prompted a large number of studies conducted on different samples of X-ray light curves in order to single out their main properties \citep{2006ApJ...642..354Z,2006ApJ...647.1213O,2007ApJ...662.1093W,2007ApJ...666.1002Z,2007ApJ...670..565L,2008ApJ...675..528L,2009ApJ...707..328L,2008ApJ...689.1161G,2009ApJ...698...43R,2009MNRAS.397.1177E}.

Analysis of the observational properties of X-ray flares \citep{2007ApJ...671.1903C,chinca10,giantflares10,2011A&A...526A..27B,2011MNRAS.410.1064M} allowed us to establish a connection with the prompt emission. In particular, the extension of the prompt lag-luminosity relation to X-ray flares strongly suggests that X-ray flares and prompt emission pulses are produced by the same mechanism \citep{giantflares10}. The nature of the underlying X-ray \emph{continuum} is still debated. It was immediately clear soon after the first set of observations by \emph{Swift}/XRT \citep{2005SSRv..120..165B} and redshift measurements that the light curves differed from a simple power-law behaviour and that some of the characteristics repeat in a systematic way  \citep[see][]{2006ApJ...642..389N,2006ApJ...642..354Z,2006ApJ...647.1213O}. After six years of data collection, we now have a much larger sample for investigating the morphology of the X-ray light curves, and a rather large subsample of redshift measurements for this analysis in the rest frame. The rest frame allows us to understand the intrinsic properties, the timescales involved, and the energetics. In fact, the parameters derived are directly related to the physics of the phenomenon, and the distributions are not affected by the distribution of redshift. This is the approach we adopt in this work.  We select light curves observed by XRT without flares, in order to avoid any kind of contamination from other emission components. When necessary we compare our results with the sample analysed in \citet{2011MNRAS.410.1064M} of GRBs with flares, providing for the first time a one-to-one comparison of light curves \emph{with} and \emph{without} flares.

The initial steep decay of the afterglow is commonly due to the tail of the prompt emission \citep{2000ApJ...541L..51K}, and the plateau phase is often connected with external shock emission with an injection of energy. However, the standard model for GRBs suffers from several problems \citep[for a review see][]{2009arXiv0911.0349L}. Here we aim at exploring alternative mechanisms to shape the light curve, such as the accretion of the stellar material left behind by the collapse of the massive star \citep{2008MNRAS.388.1729K} or the rotational energy of a newly-born magnetar \citep[see e.g.][and references therein]{2010arXiv1012.0001M}. We show indeed that an accretion model would explain the X-ray light curves well enough, as well as the properties of flares as discussed by \citet{2011MNRAS.410.1064M}. The plateau phase agrees with the assumption that new energy is injected from a spinning down neutron star into the forward shock. In particular within this framework we demonstrate that this model can account for the anticorrelation between the plateau luminosity and its end time found by \citet{2008MNRAS.391L..79D,2010ApJ...722L.215D}.

In Sect.~\ref{data_analysis} we describe the light curve extraction and the luminosity calibration procedures. In Sect.~\ref{sample} we describe the sample selection criteria, the fitting procedure and the main results about the morphology and spectral properties of the sample. In Sect.~\ref{energy} we present the analysis of the energy of the sample compared with the prompt emission energy. In Sect.~\ref{flares} we compare the present sample with the light curves with flares analysed in \citet{2011MNRAS.410.1064M}. In Sect.~\ref{discussion} we discuss our findings and in Sect.~\ref{conclusions} we draw our main conclusions. We adopt standard values of the cosmological parameters: $H_\circ=70$ km s$^{-1}$ Mpc$^{-1}$, $\Omega_M=0.27$, and $\Omega_{\Lambda}=0.73$. Errors are given at $1\, \sigma$ confidence level unless otherwise stated. Times are in rest frame unless otherwise stated.

\section{Data reduction}\label{data_analysis}

\emph{Swift}/XRT data were processed with the latest version of the \textsc{heasoft}
package available at the time of writing (v. 6.9) and
corresponding calibration files: standard filtering and screening criteria
were applied.
\emph{Swift}/XRT is designed to acquire data
using different observing modes depending on source count rates to
minimise the
presence of pile-up. However, we applied pile-up corrections when
necessary, following
the prescriptions by  \citet{2006A&A...456..917R} and \citet{2006ApJ...638..920V}.
We extracted count-rate light curves in the total XRT 0.3-10 keV energy band.
Background-subtracted, vignetting corrected light curves were then
rebinned so as
to assure a minimum signal-to-noise equals to four. When single-orbit data
were not
able to fulfil the signal-to-noise requirement, data coming from
different orbits were merged to build a unique data point.
Count-rate light curves were then converted into luminosity curves using
the spectral information derived from a time-resolved spectral analysis
where the spectral evolution
of the source, if present, is properly accounted for. We refer the reader to
\citet{2010MNRAS.402...46M} for any detail on the light curves and spectra extraction.

\begin{table}
 \begin{minipage}{85mm}
  \caption{List of the $64$ GRBs of the sample, redshift, and morphological type. The GRBs of the \emph{golden sample} are in boldface.}
  \label{tab_LC}
    \resizebox{\textwidth}{!}{
  \begin{tabular}{llllll}
  \hline
\textbf{GRB} & z$^{\dagger}$ & Type & \textbf{GRB} & z$^{\dagger}$ & Type\\
\hline
\hline
\textbf{050401} & 2.90 & Ia & 050505 &  4.27 & Ia\\
050525A &  0.606 &0& 050603 &  2.821 & 0\\
\textbf{050801} &  1.56 & Ia & 050826 &  0.287 & 0\\
\textbf{050922C} &  2.198 & Ia & \textbf{051109A} &  2.346 & II \\
051111 &  1.55 &0& \textbf{060502A} &  1.51 & II \\
060605 &  3.78 & II & 060614 &  0.125 & II \\
\textbf{060708} &  1.92 & II & 
060912A &  0.937 & 0\\
060927 &  5.47 &0& \textbf{061007} &  1.261 & 0\\
061021 &  0.3463 & II & 061126 &  1.1588 & 0\\
061222B &  3.355 & Ib & 070306 &  1.4959 & II \\
070411 &  2.954 & Ia & 070529 &  2.4996 & Ia\\
070611 &  2.04 &0& 070810A &  2.17 & Ia \\
071003 &  1.60435 &0& 
071010A &  0.98 & Ia \\
\textbf{071020} &  2.145 &0& 
071025 &  5.2 &0\\
071117 &  1.331 &0& 
\textbf{080319B} &  0.937 & 0\\
080319C &  1.95 & Ia & 
080330 &  1.51 & Ib \\
080411 &  1.03 & Ia & 
080413A &  2.433 & Ib \\
080413B &  1.1 & Ia & 
080520 &  1.545 & 0\\
080604 &  1.416 & Ib & 
\textbf{080605} &  1.6398 & Ia\\
\textbf{080707} &  1.23 & II & 
080710 &  0.845 & Ia\\
080721 &  2.591 & Ia & 
\textbf{080804} &  2.2045 & 0\\
\textbf{080905B} &  2.374 & II & 
081029 &  3.8479 & Ia\\
081118 &  2.58 & Ib & 
081121 &  2.512 & 0\\
081203A &  2.05 & II & 
\textbf{081222} & 2.77 & Ia\\
090102 &  1.547 &0& 
090313 &  3.357 & Ia\\
090323 &  3.57 &0& 
090328 &  0.736 & 0\\
090424 &  0.544 & Ia & 
090529 &  2.625 & II \\
090618 &  0.54 & II & 
090726 &  2.71 & 0\\
090814A &  0.696 & Ib & 
090926B &  1.24 & Ib \\
090927 &  1.37 & Ia & 
091003 &  0.8969 & 0\\
\textbf{091020} &  1.71 & II & 
\textbf{091109A} &  3.076 & II \\
091127 &  0.49 & Ia & 
100418A & 0.6235 & II \\
\hline
\end{tabular}	} 	 
~\\
$^{\dagger}$ for the redshift measurement, we refer to the values reported in the GRB Circular Notices.
\end{minipage}														
\end{table}

\section{Temporal and spectral properties of the X-ray light curves}\label{sample}

The sample consists of $64$ long GRBs observed by \emph{Swift}/XRT from April 2005 to April 2010. Among all the \emph{Swift} GRBs observed in this period, we selected:
\begin{itemize}
\item GRBs with redshift measurements,
\item GRBs without flaring activity in their X-ray afterglow.
\end{itemize}
For a list of the GRBs of the sample see Table~\ref{tab_LC}.

\begin{table}
 \begin{minipage}{85mm}
  \caption{Median values and standard deviations ($\sigma$) of the temporal slopes and break times for the different morphological types (see Fig.~\ref{LCtype}). The values in round brackets refer to the \emph{golden} subsample.}
  \label{tab_median}
    \resizebox{\textwidth}{!}{
  \begin{tabular}{ccccc}
  \hline
 & Type 0& Type Ia & Type Ib & Type II\\
\hline
$\left\langle \alpha_1 \right\rangle$ &  $1.29\,(1.28)$ & $0.68\,(0.80)$ & $2.9\,(-)$ & $3.3\,(3.3)$ \\
$\sigma_{\alpha1}$ & $0.27\,(0.30)$ & $0.27\,(0.38)$ & $1.3\,(-)$ & $1.2\,(0.7)$ \\
$\left\langle \alpha_2 \right\rangle$ & $-$ & $1.37\,(1.44)$ & $0.77\,(-)$ & $0.43\,(0.57)$ \\
$\sigma_{\alpha2}$ & $-$ & $0.37\,(0.27)$ & $0.37\,(-)$ & $0.32\,(0.27)$ \\
$\left\langle \alpha_3 \right\rangle$ & $-$ & $-$ & $-$ & $1.5\,(1.33)$ \\
$\sigma_{\alpha3}$ & $-$ & $-$ & $-$ & $0.4\,(0.23)$ \\
$\left\langle Log[t_{b1}/s] \right\rangle$ & $-$ & $3.25\,(3.08)$ & $2.28\,(-)$ & $1.95\,(1.85)$ \\
$\sigma_{Log[tb1/s]}$ & $-$ & $0.73\,(0.74)$ & $0.64\,(-)$ & $0.45\,(0.16)$ \\
$\left\langle Log[t_{b2}/s] \right\rangle$ & $-$ & $-$ & $-$ & $3.78\,(3.56)$ \\
$\sigma_{Log[tb2/s]}$ & $-$ & $-$ & $-$ & $0.50\,(0.31)$ \\
\hline
\end{tabular}	} 	 
\end{minipage}														
\end{table}

\subsection{Fitting procedure and classification of the light curves}\label{class}

The luminosity light curves in the rest frame were fitted adopting a simple power-law or a smoothly-joined broken power-law model (see Appendix~\ref{function})\footnote{The reported temporal index of the third segment of the smoothly double-broken power law is not the best-fitting value but the asymptotic slope (see Appendix~\ref{function}).}: a piece of software automatically identifies the smooth behaviour of the X-ray light curve of GRBs. We refer to \citet{2011MNRAS.410.1064M} for details on the fitting procedure. To reduce the number of free parameters we fix the values of the two smoothing parameters $d_1$ and $d_2$.

\begin{figure*}
\centering
\includegraphics[width=0.45 \hsize,clip]{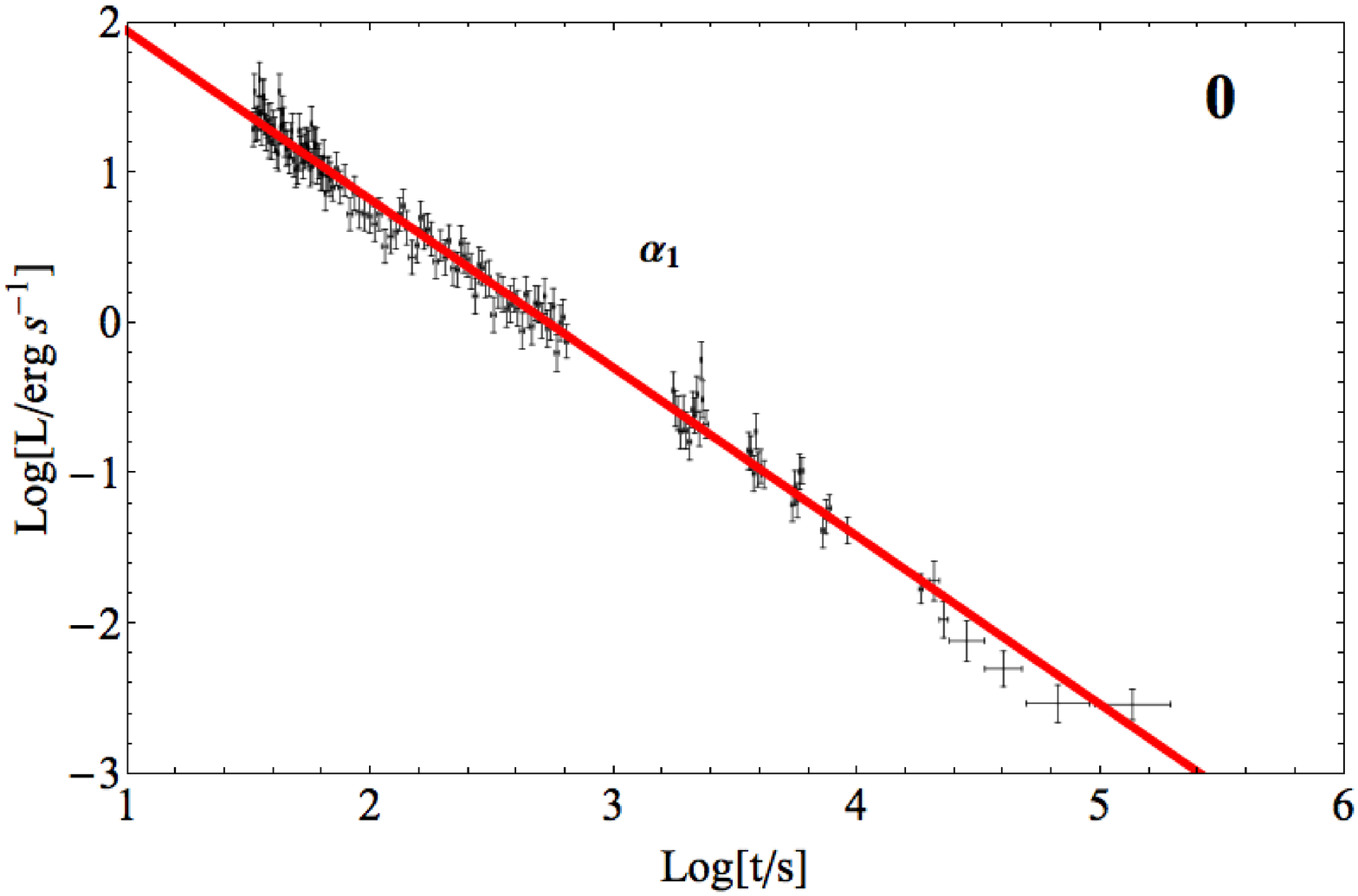}
\includegraphics[width=0.45 \hsize,clip]{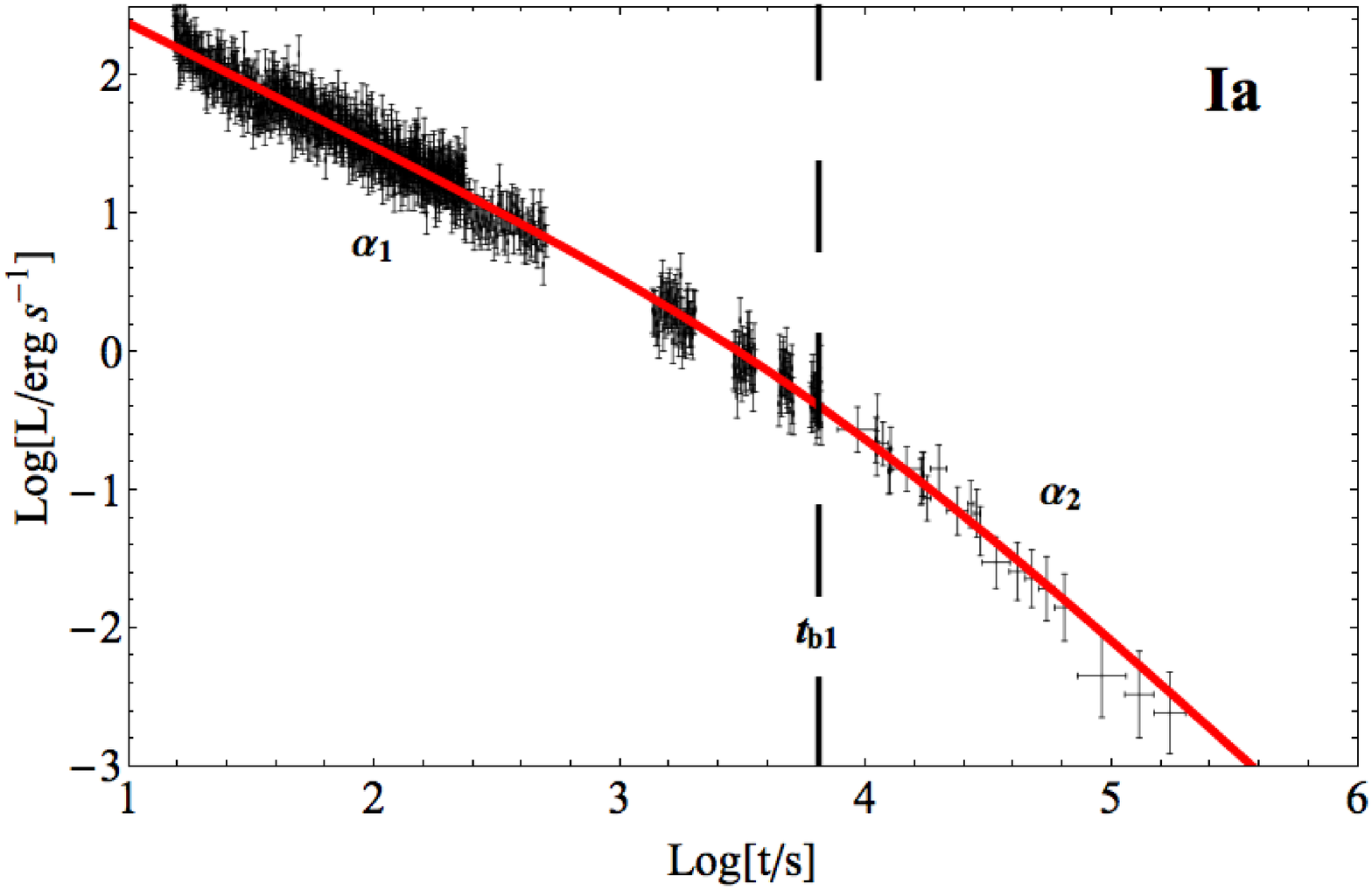}\\
\includegraphics[width=0.45 \hsize,clip]{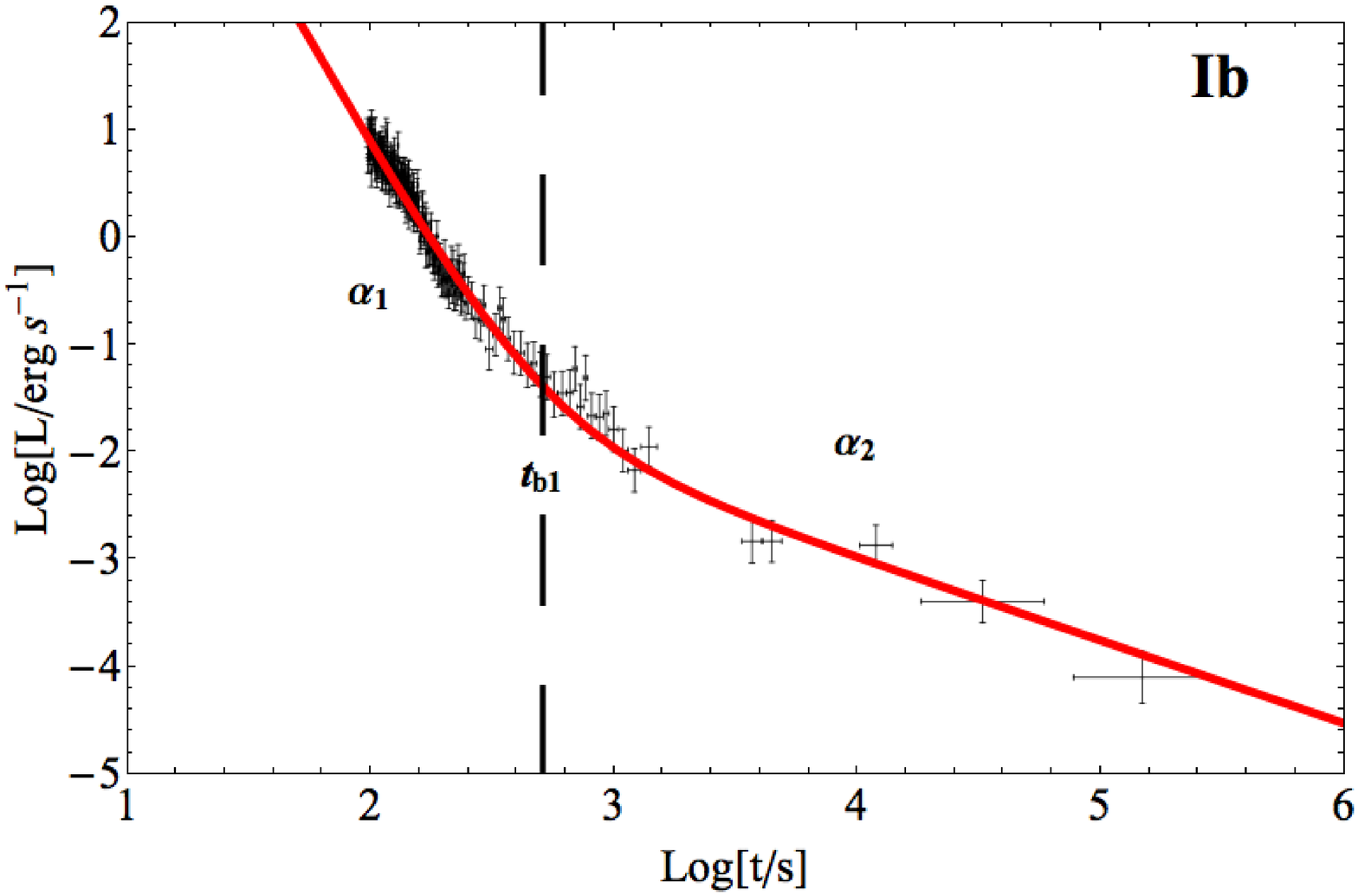}
\includegraphics[width=0.45 \hsize,clip]{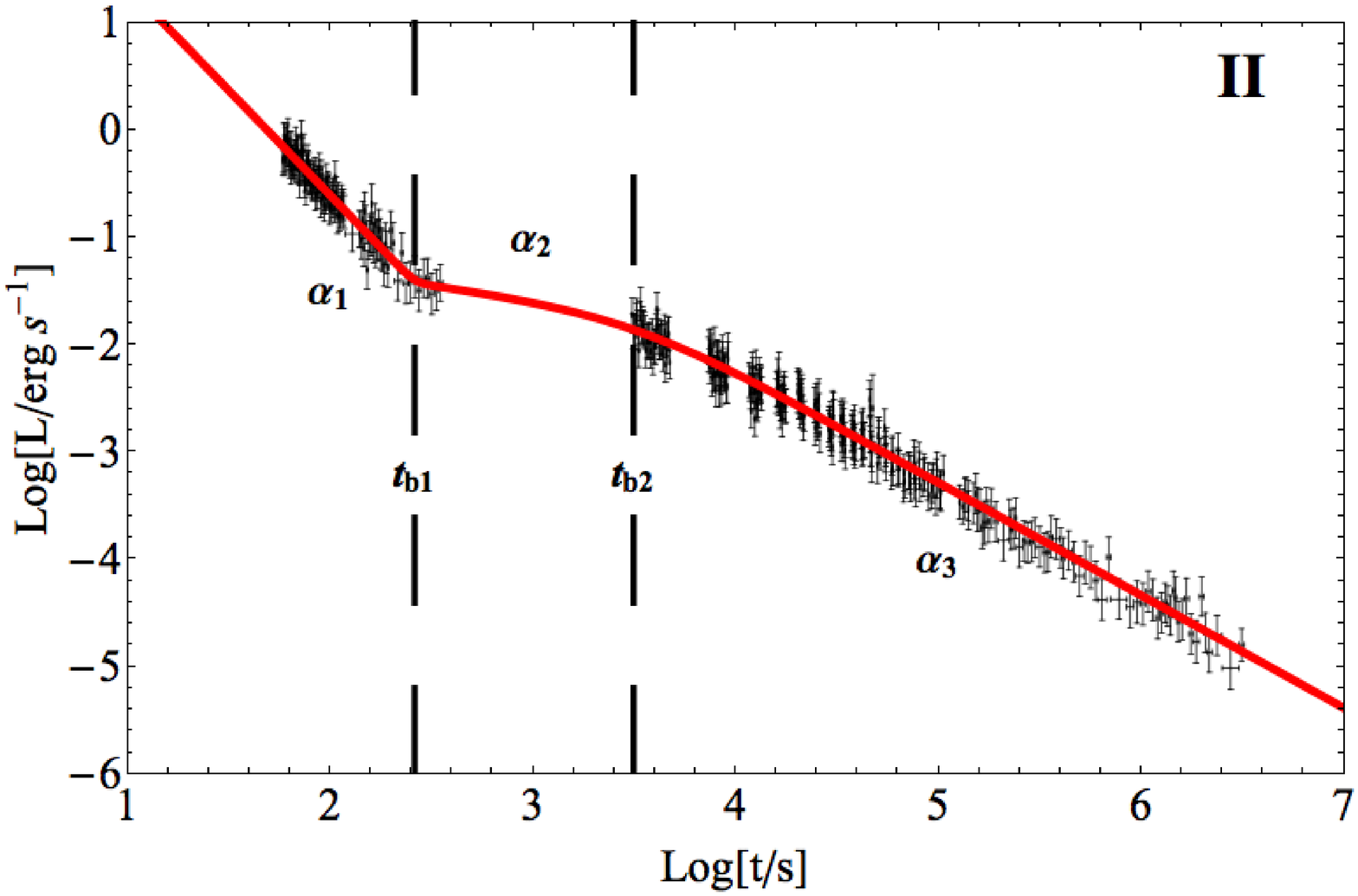}
  \caption{Examples of Type 0 (GRB080804), Type Ia (GRB081222), Type Ib (GRB090814A), and Type II (GRB061021) light curves and the respective best-fitting functions and parameters.}
   \label{LCtype}
\end{figure*}

The observed light curves can be classified on the basis of the best-fitting function\footnote{The full list of parameters is available by requesting it to grazia.bernardini@brera.inaf.it} (see Fig.~\ref{LCtype}): \textbf{Type 0} shows a single power-law decay; \textbf{Types Ia} and \textbf{Ib} show one break in the light curve and a shallow-to-steep or steep-to-shallow transition, respectively; \textbf{Type II} shows two breaks in the light curve\footnote{This classification, analogous to the one presented in \citet{2009MNRAS.397.1177E}, corresponds to types \textbf{d},\textbf{c},\textbf{b},\textbf{a}, respectively.}. A three-break model is not required to improve the best fit of any light curve of the sample. The present sample contains $22$ Type 0 light curves, $20$ Type Ia, $7$ Type Ib, and $15$ Type II. In Table~\ref{tab_median} we list the median value and standard deviation of the temporal indices of the power-law segments and the break times of the different types.

For a proper classification of the light curves, we define a \emph{golden sample} composed of $16$ GRBs that have been observed by XRT from $t_{start}\leqslant40$ s to $t_{stop}\geqslant10^5$ s in the rest frame (bold in Table~\ref{tab_LC}). The limiting start (stop) time has been chosen to be much shorter (longer) than the median first (second) break time of the Type II light curves (see Table~\ref{tab_median}). 

The presence in the golden sample of five Type Ia light curves suggests that they form a distinct class and are not simply Type II observed after the first break. In fact, in those cases the observations start time is $\gtrsim 2\,\sigma$, so earlier than the first break time of Type II, which corresponds to a probability of $\sim 5\%$. Therefore, we expect $< 1$ light curve in our Type II sample that can be included in the Type Ia golden sample. A K-S test applied to the temporal indices $\alpha_1^{Ia}$ (the first, shallow segment) of Type Ia and $\alpha_2^{II}$ (the plateau) of the Type II for the entire sample does not favour a different parent population between the two (P-value$=6\%$). Another possibility is that Type Ia light curves are Type 0 with a jet break. Although we cannot exclude this in some cases, the K-S test comparing the temporal indices $\alpha_1^{Ia}$ of Type Ia and $\alpha_1^0$ of the Type 0 for the entire sample gives a significant difference between the two (P-value$=2.2\times 10^{-6}$). Moreover, the median break time of Type Ia $\left\langle t_{b1}^{Ia}\right\rangle=1795$ s is one order of magnitude lower than the median value for jet break times of pre- and post-\emph{Swift} afterglows \citep{2009ApJ...698...43R}.

\begin{table*}
 \begin{minipage}{170mm}
  \caption{P-value of the K-S test comparing the temporal indices and break of different types of light curves in the sample.}
  \label{tab_correlazioni}
    \resizebox{\textwidth}{!}{
  \begin{tabular}{cccccccccccccc}
  \hline
& & $\alpha_1^0$ & $\alpha_1^{Ia}$ & $\alpha_2^{Ia}$ & $t_{b1}^{Ia}$ & $\alpha_1^{Ib}$ & $\alpha_2^{Ib}$ & $t_{b1}^{Ib}$ & $\alpha_1^{II}$ & $\alpha_2^{II}$ & $\alpha_3^{II}$ & $t_{b1}^{II}$ & $t_{b2}^{II}$  \\
\hline
$\alpha_1^0$ & & $1.0$ & $-$ & $-$ & $-$ & $-$ & $-$ & $-$ & $-$ & $-$ & $-$ & $-$ & $-$\\
& \\
$\alpha_1^{Ia}$ &  & $<10^{-5}$ & $1.0$ & $-$ & $-$ & $-$ & $-$ & $-$ & $-$ & $-$ & $-$ & $-$ & $-$\\
& \\
$\alpha_2^{Ia}$ &  & ${\bf0.46}$ & $<10^{-7}$ & $1.0$ & $-$ & $-$ & $-$ & $-$ & $-$ & $-$ & $-$ & $-$ & $-$\\
& \\
$t_{b1}^{Ia}$ &  & $-$ & $-$ & $-$ & $1.0$ & $-$ & $-$ & $-$ & $-$ & $-$ & $-$ & $-$ & $-$\\
& \\
$\alpha_1^{Ib}$ &  & $<10^{-4}$ & $<10^{-4}$ & $<10^{-3}$ & $-$ & $1.0$ & $-$ & $-$ & $-$ & $-$ & $-$ & $-$ & $-$\\
& \\
$\alpha_2^{Ib}$ &  & $0.007$ & ${\bf0.63}$ & $<10^{-3}$ & $-$ & $<10^{-3}$ & $1.0$ & $-$ & $-$ & $-$ & $-$ & $-$ & $-$\\
& \\
$t_{b1}^{Ib}$ &  & $-$ & $-$ & $-$ & $0.01$ & $-$ & $-$ & $1.0$ & $-$ & $-$ & $-$ & $-$ & $-$\\
& \\
$\alpha_1^{II}$ &  & $<10^{-6}$ & $<10^{-7}$ & $<10^{-5}$ & $-$ & ${\bf0.99}$ & $10^{-5}$ & $-$ & $1.0$ & $-$ & $-$ & $-$ & $-$\\
& \\
$\alpha_2^{II}$ &  & $10^{-6}$ & ${\bf0.06}$ & $<10^{-6}$ & $-$ & $10^{-5}$ & ${\bf0.49}$ & $-$ & $10^{-8}$ & $1.0$ & $-$ & $-$ & $-$\\
& \\
$\alpha_3^{II}$ &  & ${\bf0.12}$ & $10^{-6}$ & ${\bf0.36}$ & $-$ & $0.004$ & $0.004$ & $-$ & $<10^{-4}$ & $<10^{-5}$ & $1.0$ & $-$ & $-$\\
& \\
$t_{b1}^{II}$ &  & $-$ & $-$ & $-$ & $10^{-5}$ & $-$ & $-$ & ${\bf0.63}$ & $-$ & $-$ & $-$ & $1.0$ & $-$\\
& \\
$t_{b2}^{II}$ &  & $-$ & $-$ & $-$ & ${\bf0.03}$ & $-$ & $-$ & $0.002$ & $-$ & $-$ & $-$ & $<10^{-5}$ & $1.0$\\
\hline
\end{tabular}	} 	 
\end{minipage}														
\end{table*}

The golden sample does not include any Type Ib light curve. However, in three cases we have observations up to $t_{stop}\geqslant10^5$ s, thus implying that if they are Type II for which we missed the second break, this occurs remarkably later than in the Type II light curves of the sample ($t_{b2}^{II}= 6026$ s). The K-S test comparing the temporal indices $\alpha_2^{Ib}$ (the second, shallow segment) of Type Ib and $\alpha_2^{II}$ (the plateau) of Type II for the entire sample gives a $49\%$ probability that they represent the same population. The K-S test comparing the temporal indices $\alpha_1^{Ib}$ (the first, steep segment) of Type Ib and $\alpha_1^{II}$ (the steep decay) of Type II for the entire sample gives a probability of $99\%$ that they represent the same population. A connection is also found between the first break time of Type Ib $t_{b1}^{Ib}$ and of Type II $t_{b1}^{II}$ (P-value$=63\%$). Therefore, in what follows we refer to the $\alpha_1$ index of both Type Ib and Type II as the \emph{steep decay} of the GRB light curves. Likewise, we need to further investigate the nature of the \emph{shallow decay} of Types Ib $\alpha_2^{Ib}$ and Ia $\alpha_1^{Ia}$ with respect to the plateau phase $\alpha_2^{II}$ of Type II.

 Also the temporal indices $\alpha_1^{0}$ of Type 0, $\alpha_2^{Ia}$ of Type Ia, and $\alpha_3^{II}$ of Type II show significant similarities (see Table~\ref{tab_correlazioni} for a summary of the relative K-S test probabilities). We therefore group these segments as the \emph{normal decay} of the GRB light curves.

\subsection{Spectral properties of the light curves}\label{spectrum}

\begin{figure*}
\centering
\includegraphics[width=0.45 \hsize,clip]{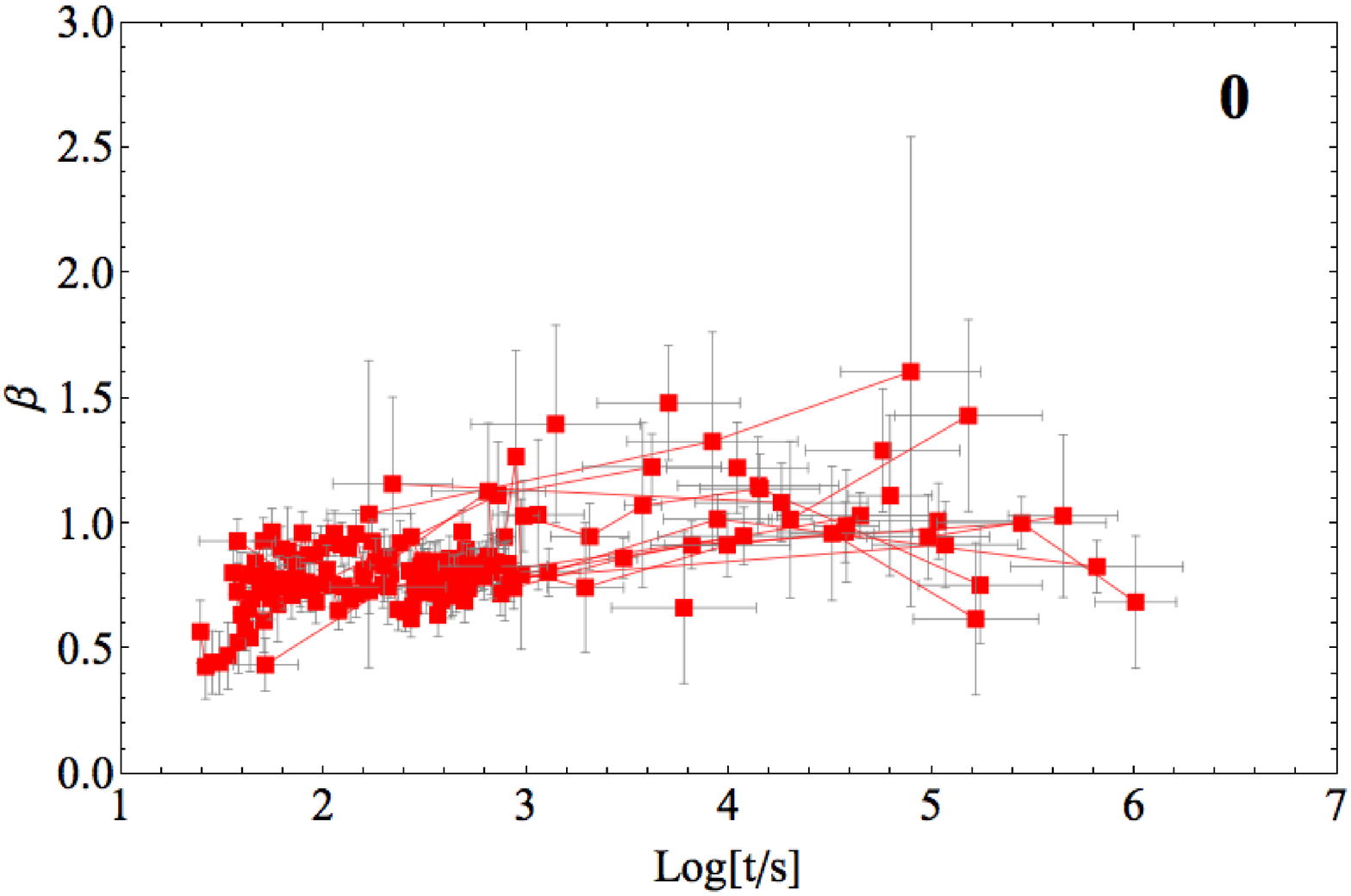}
\includegraphics[width=0.45 \hsize,clip]{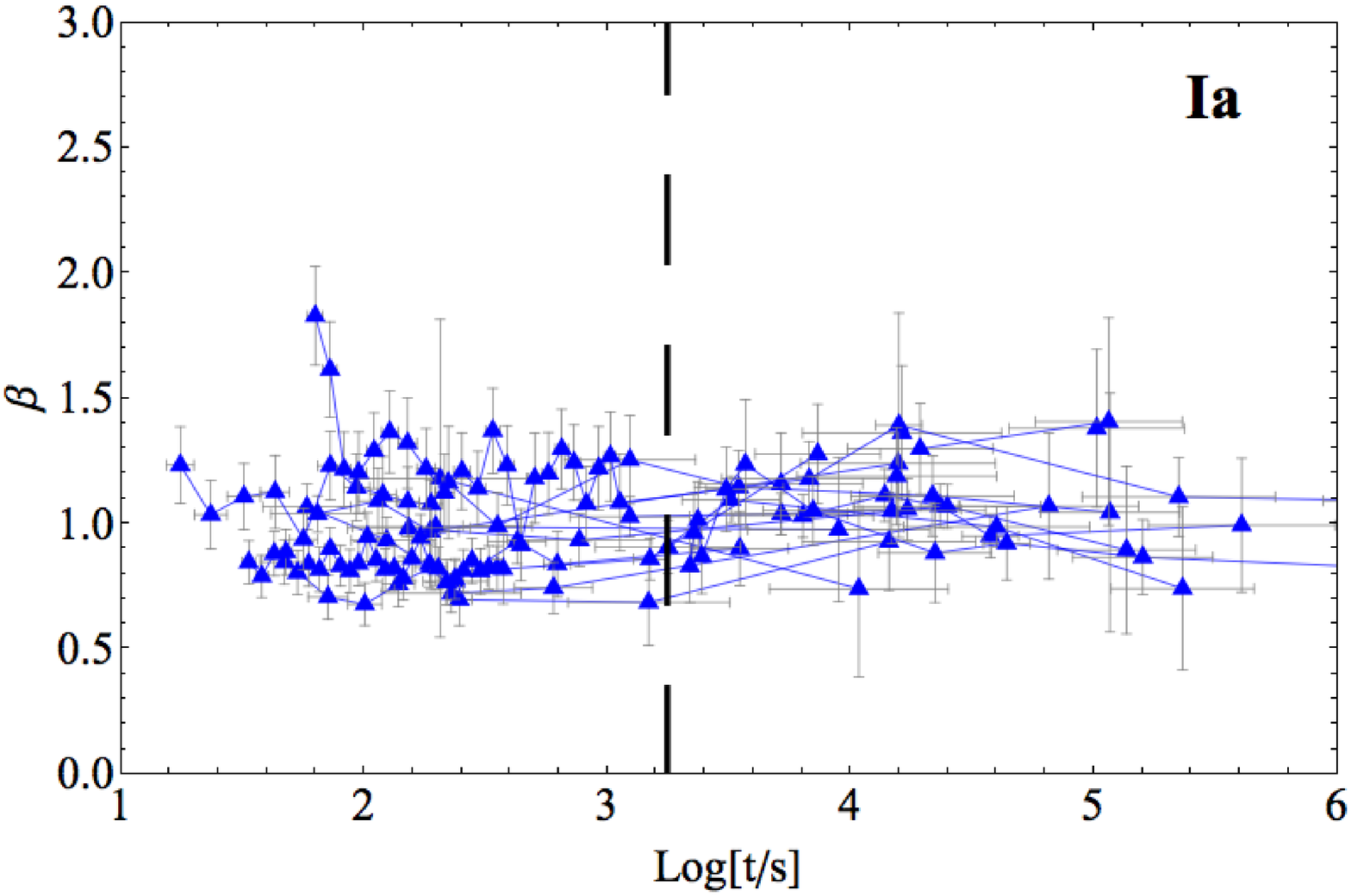}\\
\includegraphics[width=0.45 \hsize,clip]{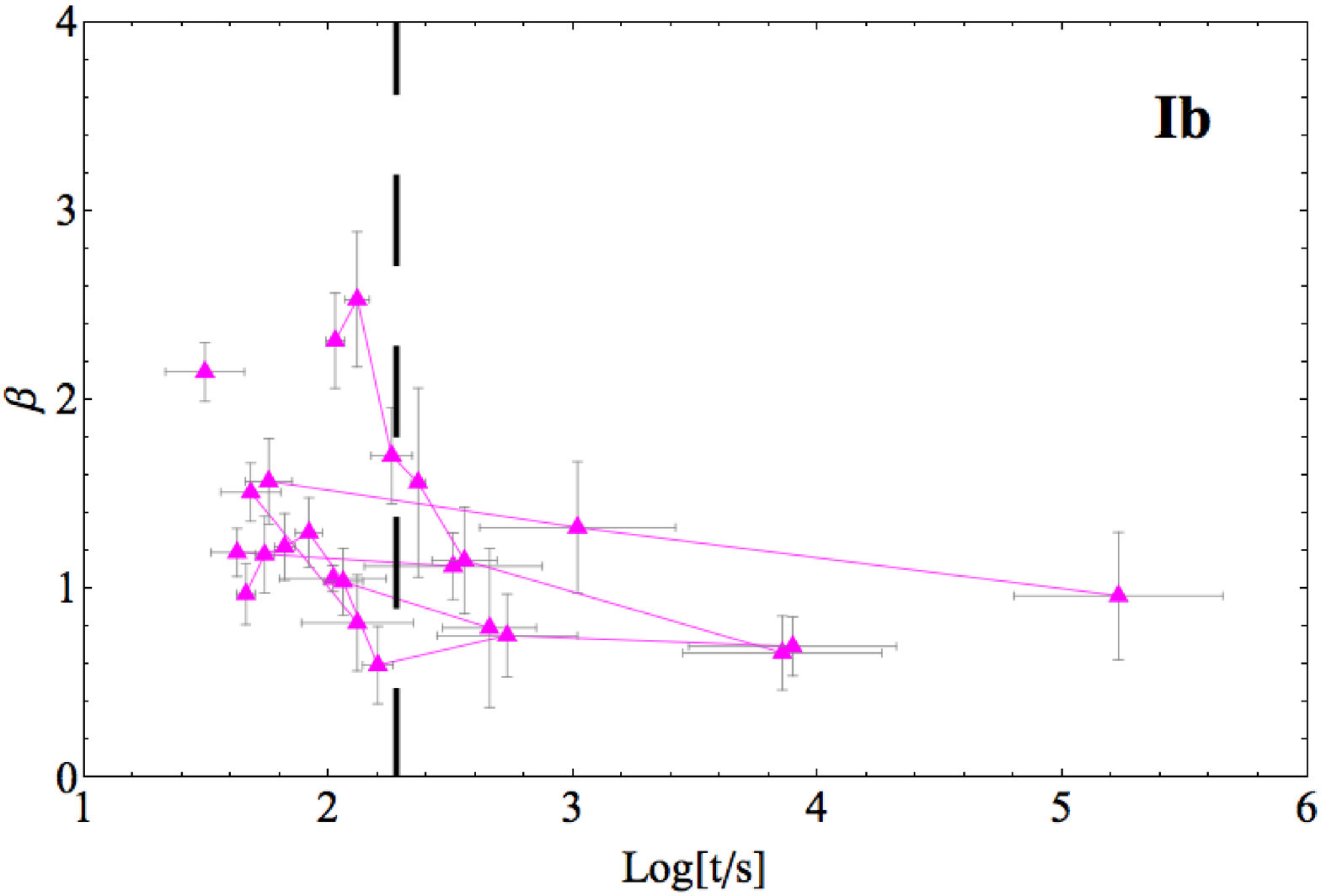}
\includegraphics[width=0.45 \hsize,clip]{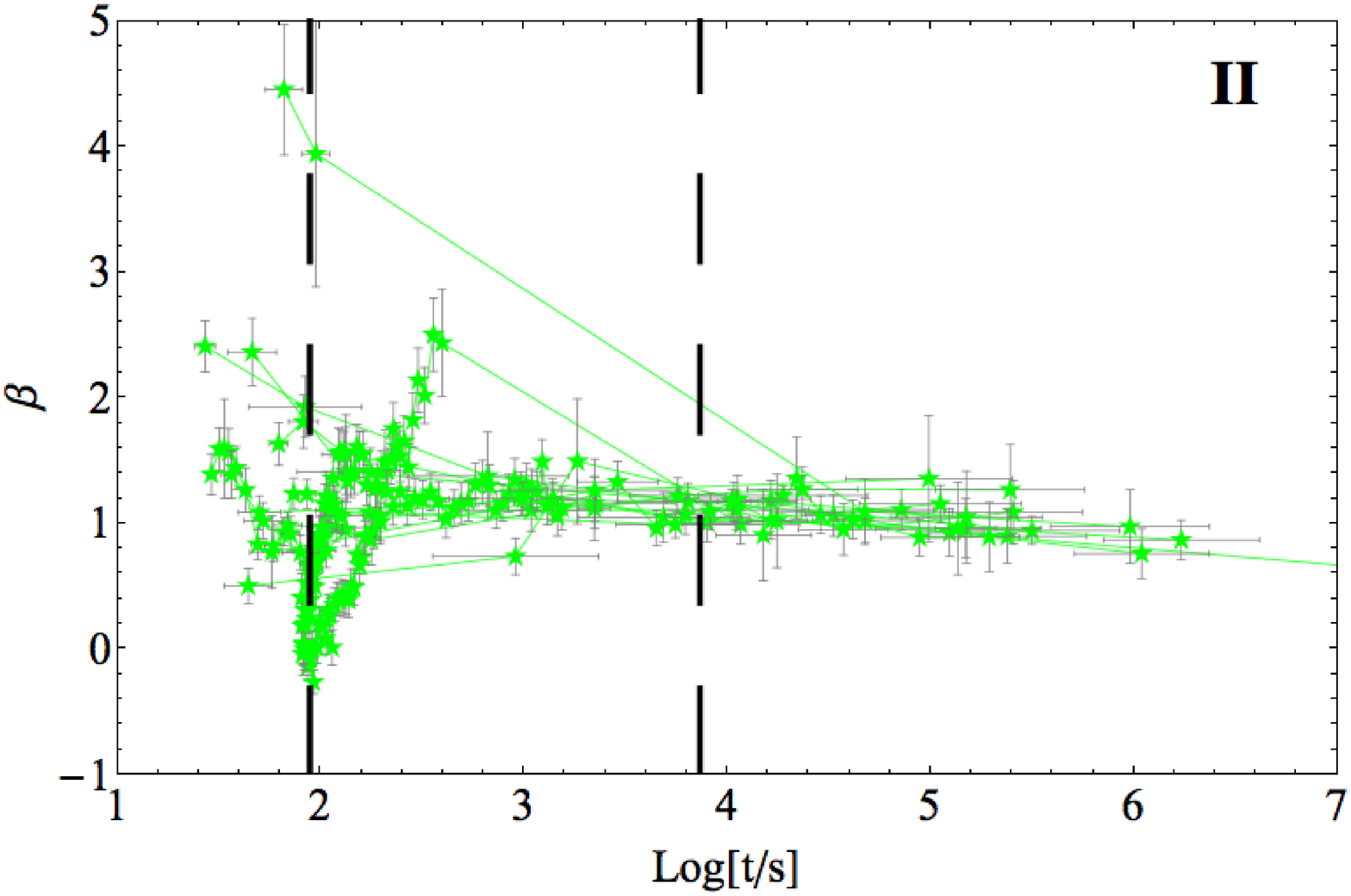}
\caption{Spectral index $\beta$ versus $t$ for the light curves belonging to Type 0 (red boxes), Type Ia (blue triangles), Type Ib (pink triangles), and Type II (green stars). The dashed black lines mark the median values of $t_{b1}$ and $t_{b2}$ for each type.}
\label{BETAtype}
\end{figure*}

The photon indices $\Gamma$ have been obtained by extracting high-quality spectra within several subsequent time intervals under the condition that $2000$ photons are included in each time interval. Then, these spectra were fit by an absorbed power law with variable $N_H$. The behaviour of the spectral index $\beta=\Gamma-1$ as a function of time for each Type of light curve is portrayed in Fig.~\ref{BETAtype}. We observe that for Type 0 light curves the spectral index is almost unchanged for the full duration of the event apart from an initial evolution. The same situation occurs for Type Ia light curves. Type Ib shows variability in the spectral indices prior to the first break time. This behaviour is analogous to Type II light curves, which show variability before the first break time and no significant variations before and after the second break times. 

The similar behaviour of the spectral indices of the steep decays of Types Ib and II enforces their association, established in the previous section. The other segments do not show any significant spectral evolution, and they are all consistent with $\tilde{\beta} \sim 1$ \citep{2007ApJ...656.1001B}. Type II light curves tend to converge to a much narrower distribution of $\beta$ than in other cases. In what follows, we consider the mean value of the spectral indices of all the spectra covering the same time interval as the spectral index of each segment of each light curve. This is an excellent approximation in all cases but the steep decays, where the average value is not always representative of the observed trend.

\section{The energy output}\label{energy}

\begin{figure*}
\centering
\includegraphics[width=0.45\hsize,clip]{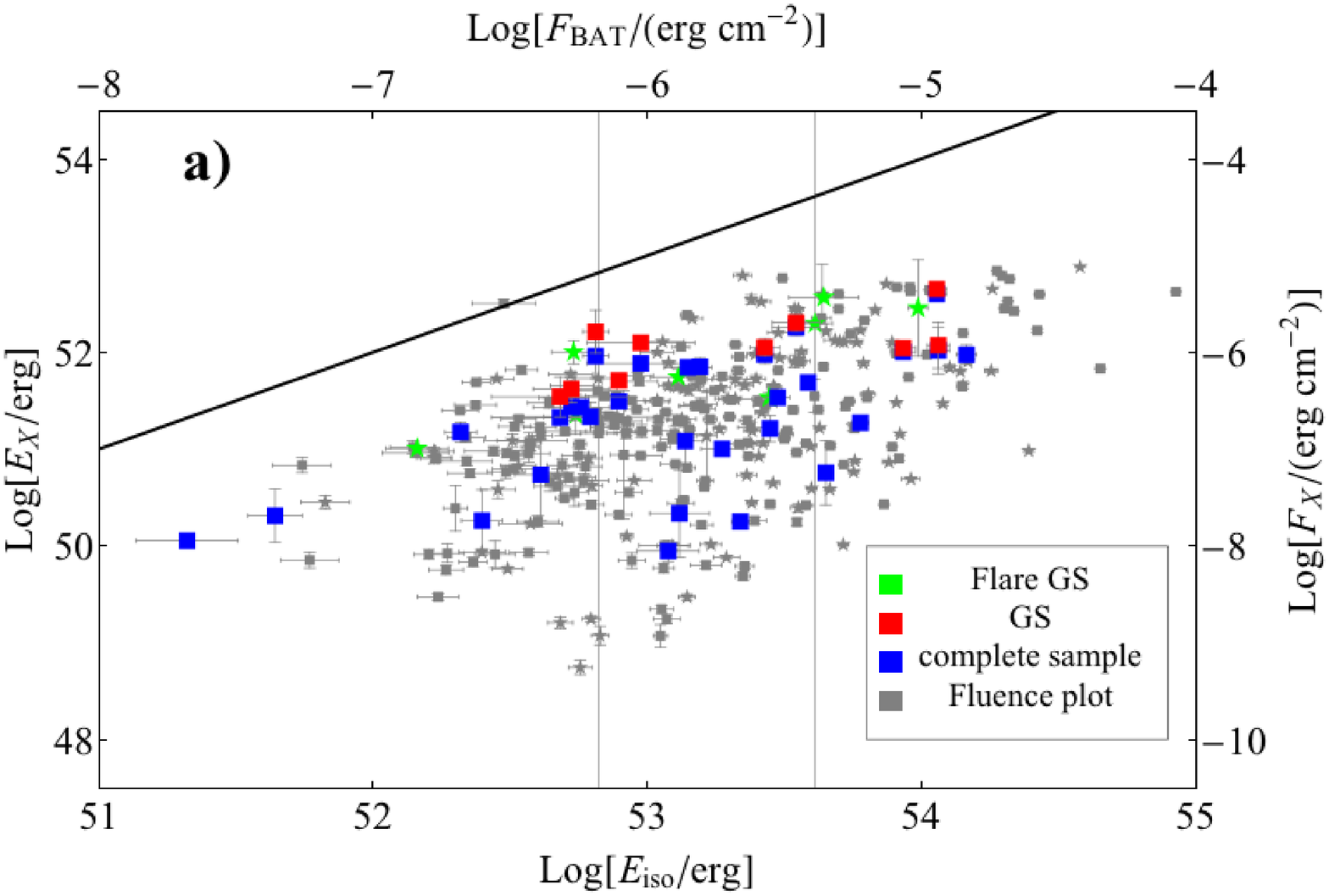}
\includegraphics[width=0.45 \hsize,clip]{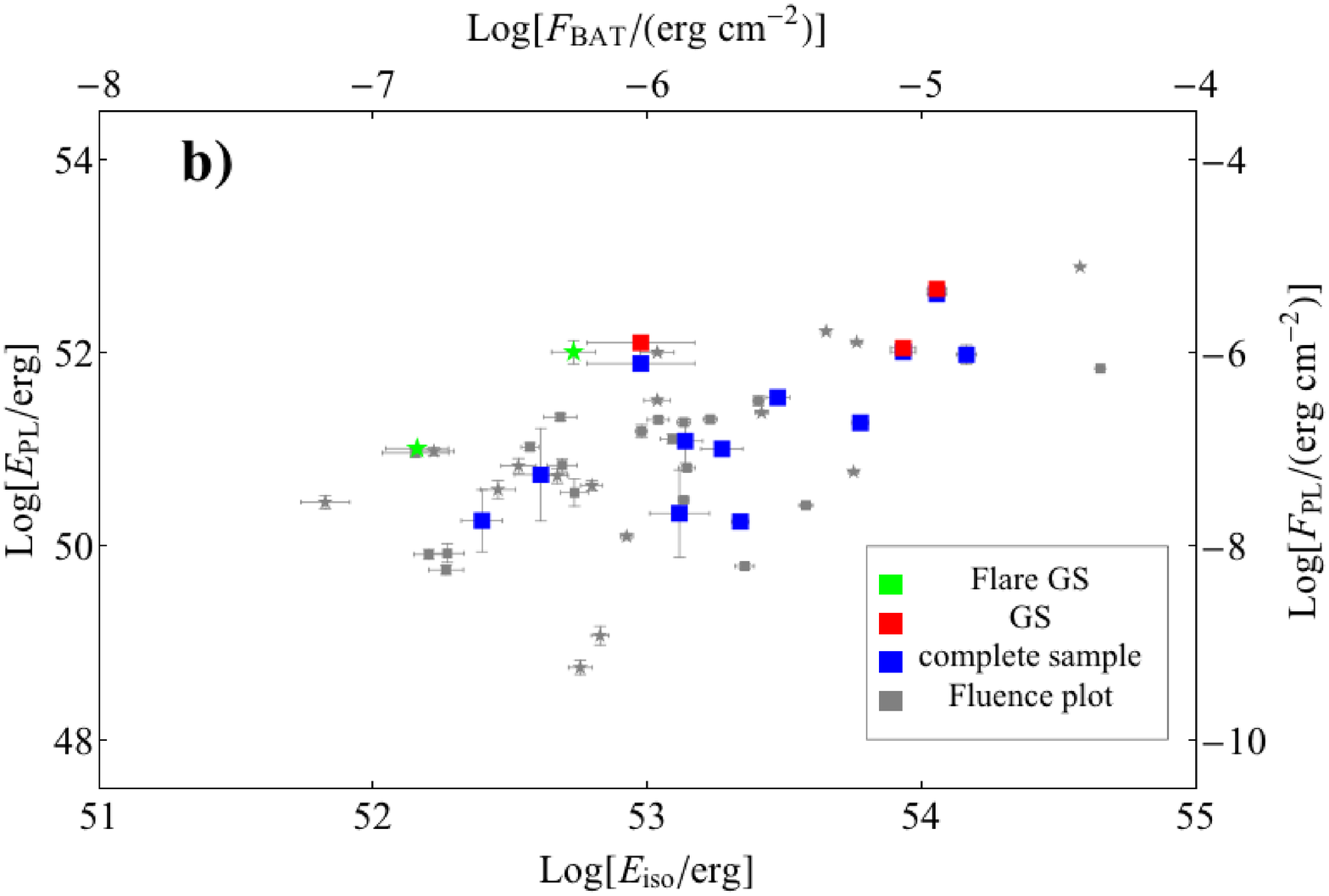}\\
\includegraphics[width=0.45\hsize,clip]{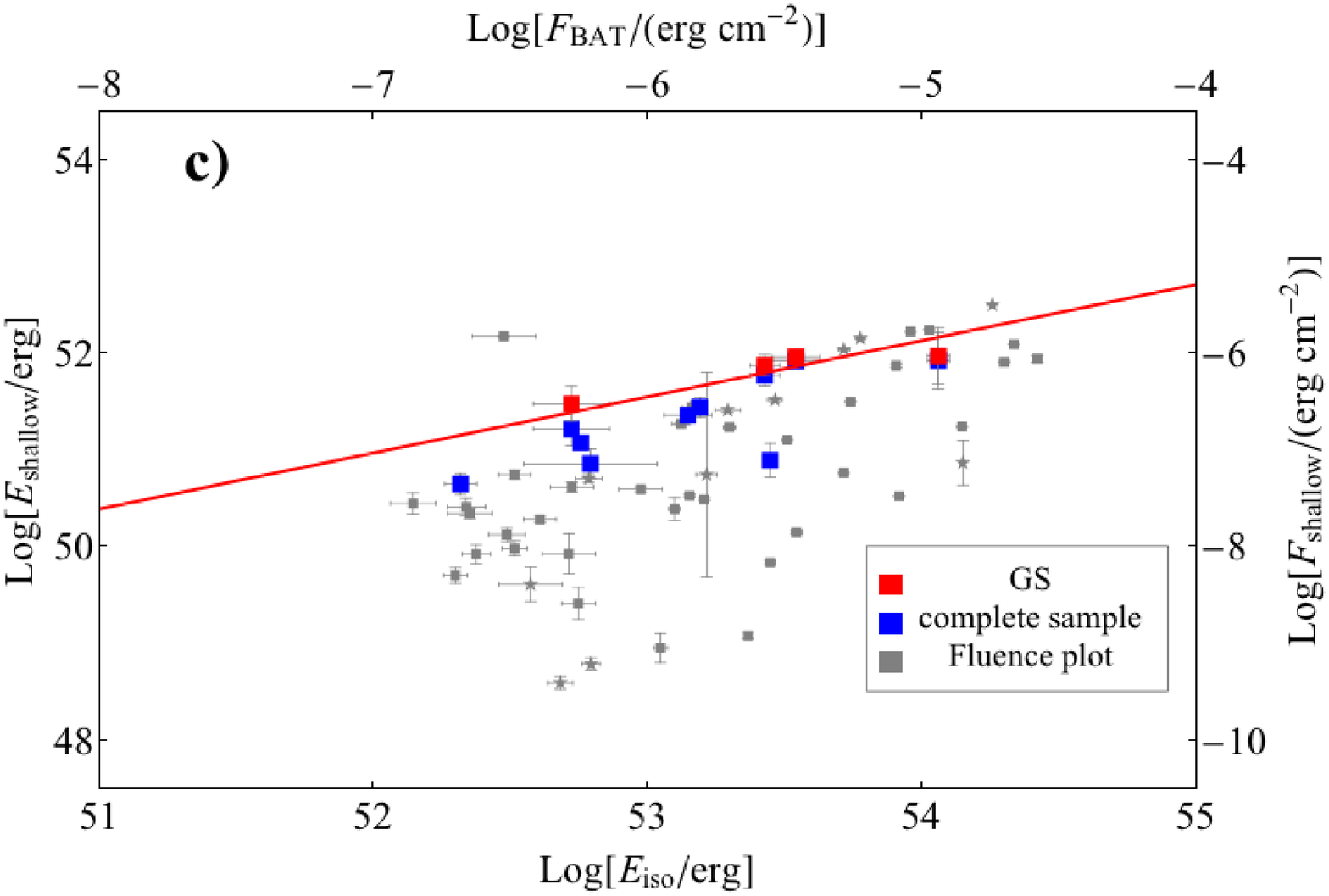}
\includegraphics[width=0.45 \hsize,clip]{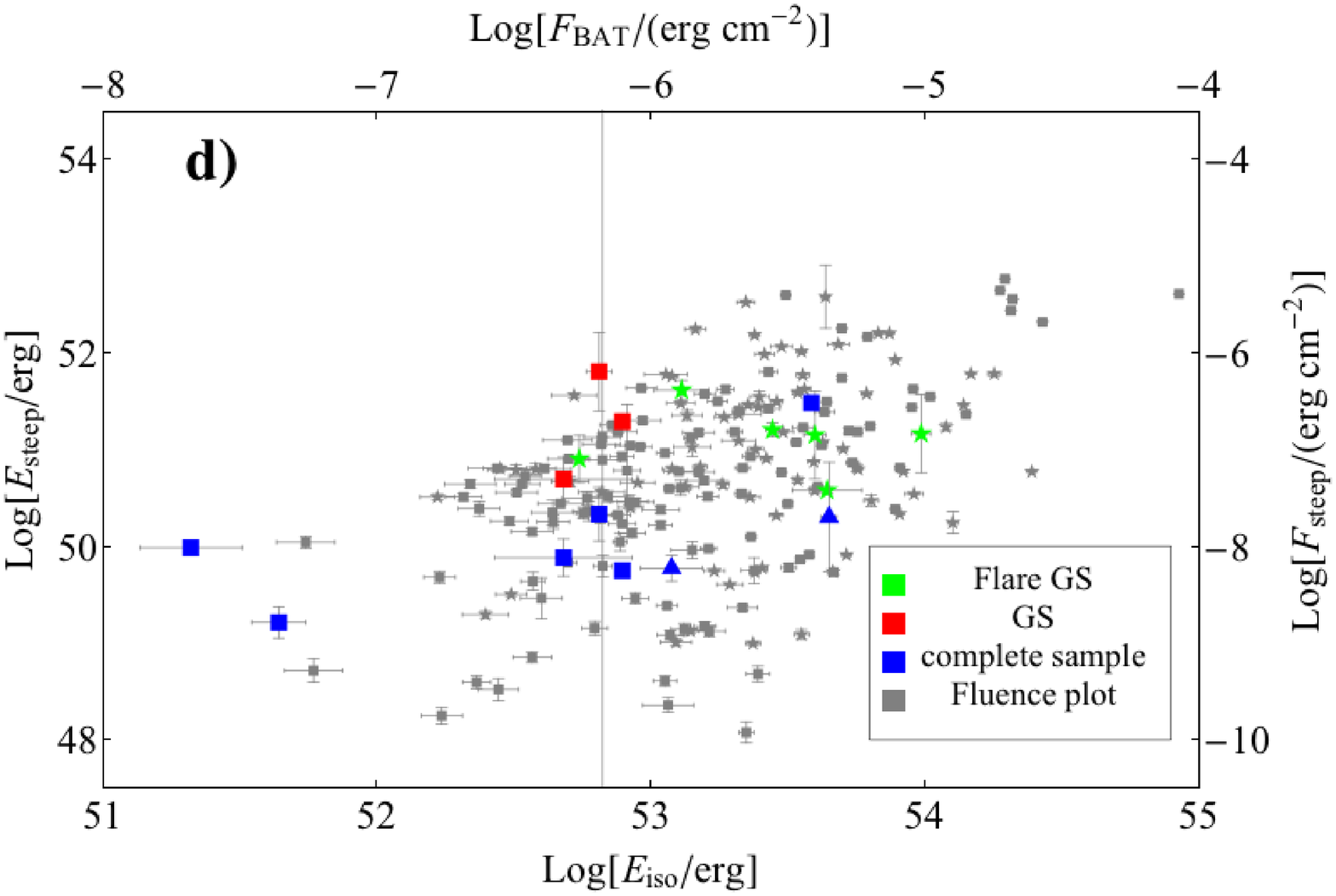}\\
\includegraphics[width=0.45 \hsize,clip]{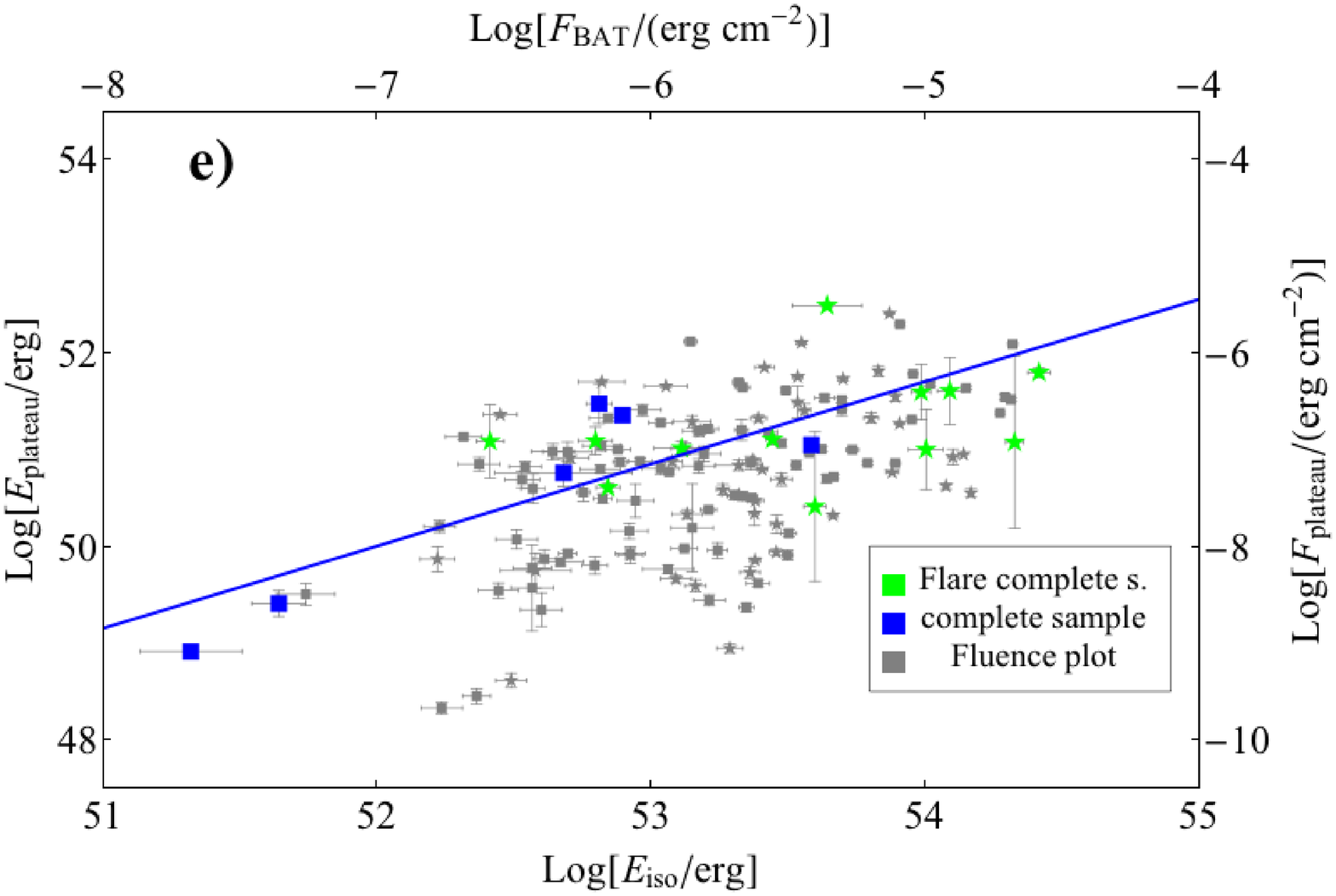}
\includegraphics[width=0.45 \hsize,clip]{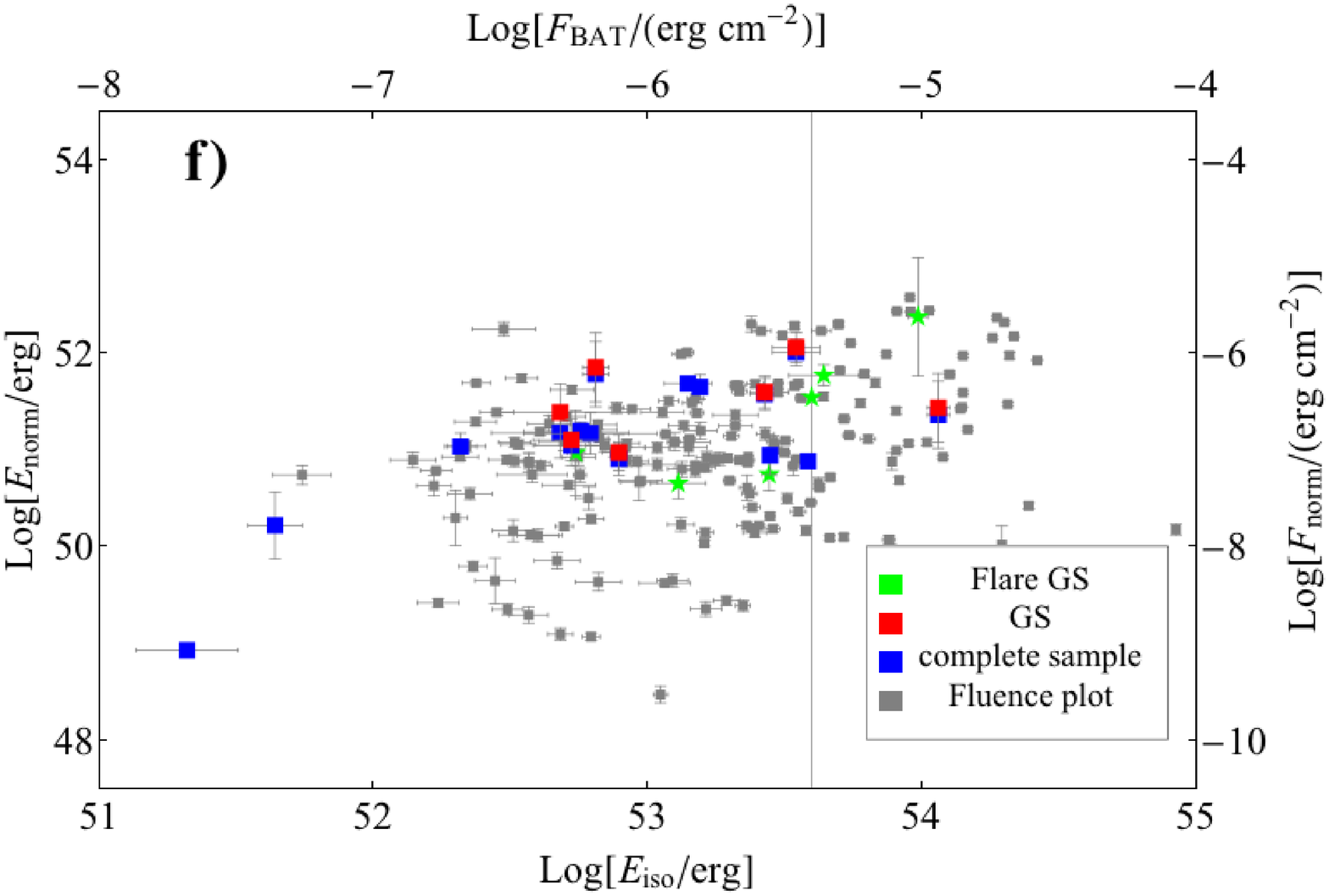}
\caption{\textbf{a)} Total energy of the X-ray light curve $E_x$ vs. prompt emission energy $E_{iso}$. The black line marks the condition $E_x=E_{iso}$. \textbf{b)} total energy of the Type 0 light curves $E_{PL}$ vs. $E_{iso}$. \textbf{c)} total energy of the first, shallow segment of Type Ia light curves $E_{shallow}$ vs. $E_{iso}$. The red line is the best-fit of the golden sample data: $Log[E_{shallow}]=(20\pm 85)+(0.6\pm1.6) Log[E_{iso}]$.  \textbf{d)} total energy of the steep decay of Type Ib and II light curves $E_{steep}$ vs. $E_{iso}$. \textbf{e)} total energy of the plateau of Type II light curves $E_{plateau}$ vs. $E_{iso}$. The blue line is the best fit of the complete sample data: $Log[E_{plateau}]=(5.8\pm 1.4)+(0.85\pm0.03) Log[E_{iso}]$.  \textbf{f)} total energy of the final normal decay of Type Ia and Type II light curves $E_{norm}$ vs. $E_{iso}$. The grey dots refer to the fluence-fluence plot of the $278$ long GRB light curves with (stars) and without (squares) flares from the XRT Catalogue by Margutti et al. (in preparation).}
\label{etype}
\end{figure*}

The energy output of the different types of X-ray light curves has been calculated as the integral of the best-fitting function over the observing time, and isotropic emission is assumed. A K-correction \citep{2001AJ....121.2879B} has been applied in order to evaluate the energies in a common rest frame energy band ($1.9-11.2$ keV). For the prompt emission isotropic energy $E_{iso}$, we refer to \citet{2008MNRAS.391..577A} and \citet{2009ApJ...704.1405K}, and to \citet{2010arXiv1012.3968N} for those GRBs observed by \emph{Fermi}/GBM whose prompt emission spectrum has been fit with a Band function. Figure~\ref{etype}.a shows the distribution of the total energy output of the X-ray light curves $E_x$ as a function of $E_{iso}$. It shows a moderate, positive correlation (with a Spearman rank $\rho=0.56$, see also \citealt{2007ApJ...662.1093W}).

We overplotted in Fig.~\ref{etype}.a the total fluence in the XRT energy band ($0.3-10$ keV) as a function of the prompt emission fluence detected by BAT ($15-150$ keV) for the complete sample of long GRBs observed by XRT until December 2010 without flares and with a complete monitoring of the X-ray light curve from before $300$ s in the observer frame until it reaches the background level. These data are provided by the XRT Catalogue (Margutti et al., in preparation; see also \citealt{2011arXiv1107.2870Z}). The existence of a positive correlation between the X-ray and prompt emission is also apparent in the observer frame ($\rho=0.51$).

The X-ray energy that we computed for the present sample is a lower limit of the total, since we only consider the emission during the observing time in a limited energy band. To partially overcome this limit, we consider the light curves belonging to the golden sample separately, and we integrate those curves between the end of the prompt emission (the rest-frame $T_{90}$) and $10^7$ s. The positive correlation between the prompt emission and the X-ray energy for this subsample is enhanced ($\rho=0.59$, see Fig.~\ref{etype}.a). 

We now consider each light curve type separately. Type 0 light curve energy $E_{PL}$ shows a moderate correlation with the prompt emission energy, but the paucity of Type 0 light curves in the golden sample does not allow us to be more conclusive (see Fig.~\ref{etype}.b). The same situation is found for the steep decay of Type II and Ib energy $E_{steep}$ (see Fig.~\ref{etype}.d) and for the normal decay of Type II and Ia energy $E_{norm}$ (see Fig.~\ref{etype}.f). When we consider the corresponding fluences as a function of the $15-150$ keV fluence of the prompt emission for the sample defined above\footnote{The light curves of the XRT Catalogue are fitted and classified with the same procedure as adopted in this work, therefore we are comparing equivalent quantities for each part of the light curves \citep[see][]{2011arXiv1107.2870Z}.}, we do not find any trend toward Type 0 light curves ($\rho=0.35$) or toward the normal decay ($\rho=0.33$), and a moderate correlation for the steep decay ($\rho=0.49$).

The energy output of the plateau phase $E_{plateau}$ in Type II light curves is positively correlated with $E_{iso}$ (see Fig.~\ref{etype}.e). In this case we can consider the entire sample (the initial and end time of the plateau are always known), and we find a linear dependence: $E_{plateau}\propto E_{iso}$, with $E_{plateau}/E_{iso}\sim 2\%$, which is roughly the same order of magnitude than $E_{flare}/E_{iso}$ for early-time flares \citep{chinca10,2011A&A...526A..27B}. A correlation is also found between the prompt emission energy and the initial shallow decay of Type Ia energy $E_{shallow}$ (see Fig.~\ref{etype}.c). If we consider only the golden sample, we find that this dependence is almost linear: $E_{shallow}\propto E_{iso}$, with $E_{shallow}/E_{iso}\sim 3\%$. Both these correlations are also apparent in the corresponding fluence-fluence plot ($\rho=0.58$ and $\rho=0.56$, respectively). We have to caution that a prompt monitoring of the light curve is necessary in order to distinguish Types Ia and II. Therefore, the fluence-fluence plot for the shallow decay can contain spurious events due to a misclassification.

\section{X-ray light curves with and without flares}\label{flares}

The selection of X-ray light curves without flares allows us to investigate whether the X-ray continuum knows about the presence of flares. For this, we compare our results with the corresponding values of the sample considered in \citet{2011MNRAS.410.1064M} of $44$ GRBs that exhibit flaring activity. The major advantage is that the fitting procedure and the functions used for the continuum underlying the X-ray light curve of that sample are the same\footnote{We recall, however, that for the flare sample the luminosity is in the common rest-frame energy band $2.2-14.4$ keV.}.

If we classify the light curves with flares (FLC) on the basis of their best-fitting function as in Sect.~\ref{class} we find that the sample of \citet{2011MNRAS.410.1064M} contains a majority of Type II light curves ($72\%$), $14\%$ of Type 0, and $7\%$ of Types Ia and Ib. This repartition is different from what we found in our sample ($34\%$ of Type 0, $31\%$ of Type Ia, $11\%$ of Type Ib, and only $23\%$ of Type II, see also \citealt{giantflares10}). 

We produced a population of ``fake'' flares on the basis of the relations found in \citet{chinca10} ($w=0.2\,t_{pk}+10$; $t_{dec}\sim2\,t_{rise}$; $L_{pk}\propto t_{pk}^{-2.7}$; $E_{flares}\sim10^{51}$ erg) with $40$ s $\leqslant t_{pk}\leqslant 500$ s, and we superimposed these flares to the light curves of our golden sample. We find that a ``standard'' flaring activity would have been detected above $1 \sigma$ in $80\%$ of the golden sample light curves. In particular, these flares could have been detected in all the five Type Ia light curves for $t \gtrsim 100$ s. Moreover, if we compute the average luminosity of the light curves of the golden sample, we find that it has a shallower decay ($\left\langle L\right\rangle\propto t^{-2}$) but is even dimmer than the average continuum obtained in \citet{2011MNRAS.410.1064M} for $t \lesssim 200$ s. This excludes a luminosity bias preventing detection of early time flares in the light curves of our golden sample.

\begin{figure}
\includegraphics[width=\hsize,clip]{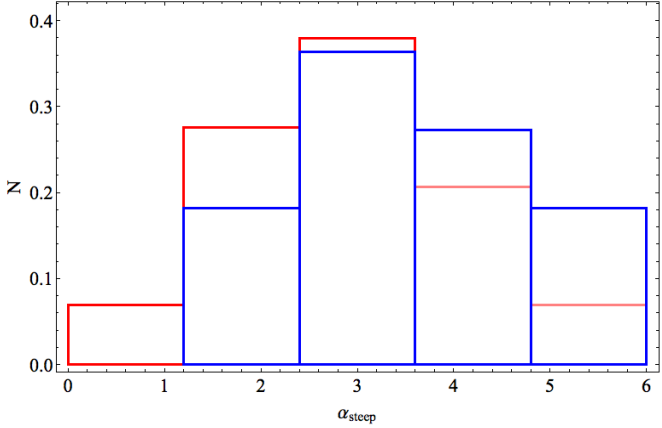}
\caption{Probability distribution of the steep decay power-law index $\alpha_{steep}$ for the LC (blue lines) compared with the FLC sample (red line).}
\label{alpha}
\end{figure}

In Fig.~\ref{alpha} we compare the distribution of the steep decay power-law index of Type II and Ib light curves of the present sample (LC) with the corresponding value found for the FLC. We did not include those cases in which the prompt emission still observed in the X-rays (as GRB060510B). The FLC power-law index distribution has a median $\left\langle \alpha^{FLC}\right\rangle=2.7$ ($\sigma^{LC}=1.3$), while the LC sample has $\left\langle \alpha^{FLC}\right\rangle=3.2$ ($\sigma^{LC}=1.2$). The K-S test comparing the temporal indices $\alpha^{LC}$ and $\alpha^{FLC}$ gives a probability of $9\%$ that they belong to the same population. \citet{2011MNRAS.410.1064M} show that light curves with multiple flares tend to be shallower than those with a single flare. We then select the FLC with a single flare that are expected to be more similar to LCs and compare their power-law indices with the LC distribution. The two distributions are more similar, with a median of the single FLC distribution $\left\langle \alpha_{sing}^{FLC}\right\rangle=3.0$ ($\sigma_{sing}^{FLC}=1.0$). The K-S test comparing the temporal indices $\alpha^{LC}$ and $\alpha_{sing}^{FLC}$ results in an increasing probability that they belong to the same population ($27\%$).

We aim at investigating now the late-time behaviour of the entire sample. For this purpose we computed the average light curve for the entire LC sample. After $1000$ s we observe that $\left\langle L\right\rangle\propto t^{-1.2}$, a trend similar to the detection threshold found for the average flare light curve in \citet{2011MNRAS.410.1064M}. If we compare the average luminosity of the LC sample with the one of the continuum underlying the late-time flares (LFLC) of \citet{2011A&A...526A..27B}, we find that the temporal behaviour is very similar but the present sample is brighter, with $L^{LFLC}/L^{LC}\sim 20\%$.

We turn now to the energy output of the X-ray light curves with flares. We restrict our analysis to a \emph{flare golden sample} (FGS) of $17$ light curves, defined with the same criterion as was introduced in Sect.~\ref{sample}, and we calculate the energy of the underlying continuum in the common rest frame energy band ($1.9-11.2$ keV) as in Sect.~\ref{energy}. We find that the distributions of both the X-ray energy and the isotropic prompt emission energy are similar for the two samples. A correlation between the $E_x$ and $E_{iso}$ is present ($\rho=0.70$, see Fig.~\ref{etype}), as well as between the total fluence in the XRT energy band ($0.3-10$ keV) and the prompt emission fluence detected by BAT ($15-150$ keV) for the complete sample of long GRBs with and without flares from the XRT Catalogue by Margutti et al. (in preparation) with a complete light curve in the sense described in the previous section ($\rho=0.51$). The correlation between $E_{plateau}$ and $E_{iso}$ is enhanced including the FGS data (see Fig.~\ref{etype}.e). The results of previous section are confirmed by adding the light curves with flares also in the fluence-fluence plot for the plateau phase ($\rho=0.50$) and the shallow decay ($\rho=0.63$). The other fluence-fluence relations portrayed in Fig.~\ref{etype} are not improved significantly with the exception of the Type 0 versus prompt emission ($\rho=0.50$).

\section{Discussion}\label{discussion}

The analysis of the rest-frame X-ray light curves of our sample shows that they can be grouped into four morphological types. We found that the steep decay of Types Ib and II can be grouped together, as well, as the shallow decays in Types Ia, Ib, and II, and normal decays in Types 0, Ia, and II. Spectral evolution is associated with the steep decay, while no spectral evolution is found during shallow and normal decay. We demonstrated that

\begin{itemize}
\item a quasi-linear correlation exists between the total isotropic energy emitted during the prompt phase $E_{iso}$ and the energy released during the X-ray shallow decay phase of GRBs with and without flares. This correlation is also apparent in the observer frame of a larger sample;
\item the predominance of Type II with flares with respect to the present sample indicates that the probability of flare occurrence is different depending on the light curve morphology;
\item the energetic of the X-ray continuum with and without flares is similar. Light curves without flares tend to be steeper than those with flares.
\end{itemize}

\subsection{The steep decay}\label{steepdecay}

The widespread consensus is that the initial steep decay observed in Types Ib and II is the tail of the prompt emission. If the central engine stops its activity at the end of the prompt emission, the decline in the X-ray light curve is imposed by the high-latitude emission \citep[HLE, $t^{-2-\beta}$;][]{2000ApJ...541L..51K}. This rapid decline is possible only if $\theta_j>\Gamma^{-1}$, where $\theta_j$ is the jet opening angle and $\Gamma$ the Lorentz gamma factor of the ejecta. If $\theta_j<\Gamma^{-1}$, the light curve morphology is determined by the emission of ejecta that cool via adiabatic expansion (AE) \citep[AE;][]{2009MNRAS.395..955B}. Variations in the decay rate are expected for structured jets \citep{2002ApJ...571..876Z,2006MNRAS.369L...5L}. 

\begin{figure}
\includegraphics[width=\hsize,clip]{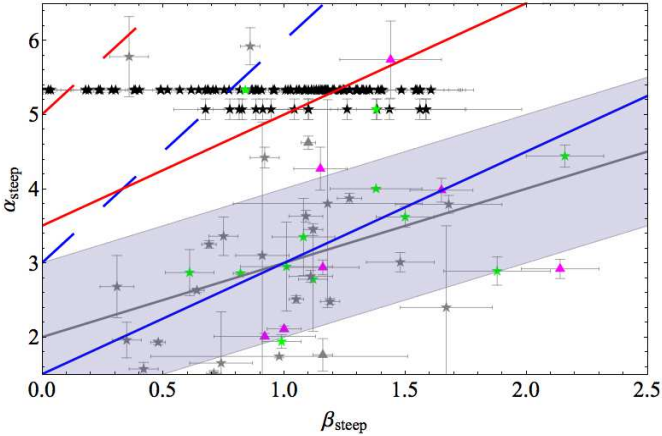}
\caption{Temporal decay index $\alpha_{steep}$ vs. spectral index $\beta_{steep}$ of the first segments of Type II (green stars) and Type Ib (pink triangles) light curves of the present sample. Grey symbols refer to the sample of \citet{2011MNRAS.410.1064M}. The black stars track the positions for the different $\beta(t)$ of GRB090618 and GRB081203A. The grey solid line corresponds to the expectation for the HLE: $\alpha=2+\beta$. The shaded area corresponds to the expectation for the structured jet ($-3-\beta\leqslant\alpha\leqslant-1-\beta$). The blue lines correspond to the expectations for the synchrotron emission from AE ejecta: $\alpha=1.5\beta+1.5$ for thick ejecta (solid line) and $\alpha=3\beta+3$ for thin ejecta (dashed line). The red lines correspond to the expectations for the SSC emission from AE ejecta: $\alpha=1.5\beta+3.5$ for thick ejecta (solid line) and $\alpha=3\beta+5$ for thin ejecta (dashed line).}
\label{absteep}
\end{figure}

In Fig.~\ref{absteep} we show that $32\%$ of the sample agrees with HLE within $1\sigma$, $32\%$ with AE, and $79\%$ are within the ranges imposed by the structured jet. GRB090618, GRB081203A, and GRB081118 are not compatible with any case. If we include Type Ib and II GRBs of the sample analysed in \citet{2011MNRAS.410.1064M}, we find that $82\%$ are consistent with one of the above cases.

The simplest scenario predicts that no spectral evolution should be observed during this stage(\citealt{2007ApJ...666.1002Z,2009MNRAS.395..955B}; see, however, \citealt{2007ApJ...663..407B}). Figure~\ref{BETAtype} clearly shows that this requirement is never met for the Type II and Ib light curves of our sample. GRB090618 and GRB081203A are clear examples: we show in Fig.~\ref{absteep} the different positions that GRBs assume during the steep decay in the $\alpha_{steep}-\beta_{steep}$ plane due to the spectral evolution.

The spectral evolution of the early steep decay and other arguments \citep[see e.g.][]{2009MNRAS.395..955B} lead us to believe that the observed early X-ray decay of GRBs is produced by the rapidly declining continued activity of the central engine, if we assume that there is a one-to-one correspondence between the temporal behaviour of the central engine activity and the observed emission \citep{2009MNRAS.395..955B}. This is the case for a continuous accretion onto the central object, as proposed by \citet{2008MNRAS.388.1729K}. The decay indices ($\left\langle \alpha_{steep}\right\rangle = 3.15$) and spectral evolution we observe in our sample support this hypothesis. In fact, the predicted light curves decay as $L\propto t^{-3}$ or steeper when the density in the outermost layer of the progenitor star has a sharp density profile, or $L\sim t^{-2}$ for a higher rotation rate of the progenitor star core (see Figs.~4 and 5 in \citealt{2008MNRAS.388.1729K}, see also \citealt{2010ApJ...713..800L} for the results of hydrodynamic simulations).

Within this framework, flares could be powered by instabilities affecting the physical source of energy, which creates the steep decay (\citealt{2008MNRAS.388.1729K}, see also the results from \citealt{2011MNRAS.410.1064M}). This is a plausible explanation for the higher occurrence probability associated with the steep decay. Therefore this scenario also has the merit of accounting for the connection steep decay/flaring activity.

\subsection{The plateau phase}\label{sectplat}

The intermediate shallow decaying phase observed in Type II light curves is interpreted as an injection of energy into the forward shock \citep[see e.g.][and references therein]{2006ApJ...642..354Z}. The absence of significant spectral evolution during this stage agrees with the expectations from forward shock emission (see Fig.~\ref{BETAtype}). Alternative possibilities have been discussed by \citet{2006MNRAS.369..197F}, including a small fraction of electrons being accelerated, evolving shock parameters, low non-constant density of the interstellar medium, and a magnetised outflow \citep[see also][]{2003astro.ph.12347L}. These authors conclude that the most viable scenario is the energy injection model, although it cannot self-consistently account for both the X-ray and optical afterglows in all cases \citep[see e.g.][]{2011MNRAS.414.3537P}.

\subsubsection{Energy injection from spinning-down neutron star}\label{magnetar}

A possible source of energy injection is the power emitted by a spinning-down newly-born magnetar \citep{1998A&A...333L..87D,2001ApJ...552L..35Z,2009ApJ...702.1171C} that refreshes the forward shock (see \citealt{2011A&A...526A.121D} for an analytic treatment of this problem). We applied the solution of \citet{2011A&A...526A.121D} to our sample (details are given in Appendix~\ref{app_magnetar}), and we find that it fits the observed light curves well enough. This is independent of the presence of flares (see Table~\ref{tab_magnetar}), provided that the light curve's normal decay agrees with the forward shock model predictions\footnote{The applicability of the magnetar model to GRBs displaying a very steep decay after the plateau inconsistent with the forward shock emission has been investigated by \citet{2010MNRAS.402..705L}.}.

\begin{figure*}
\includegraphics[width=\hsize,clip]{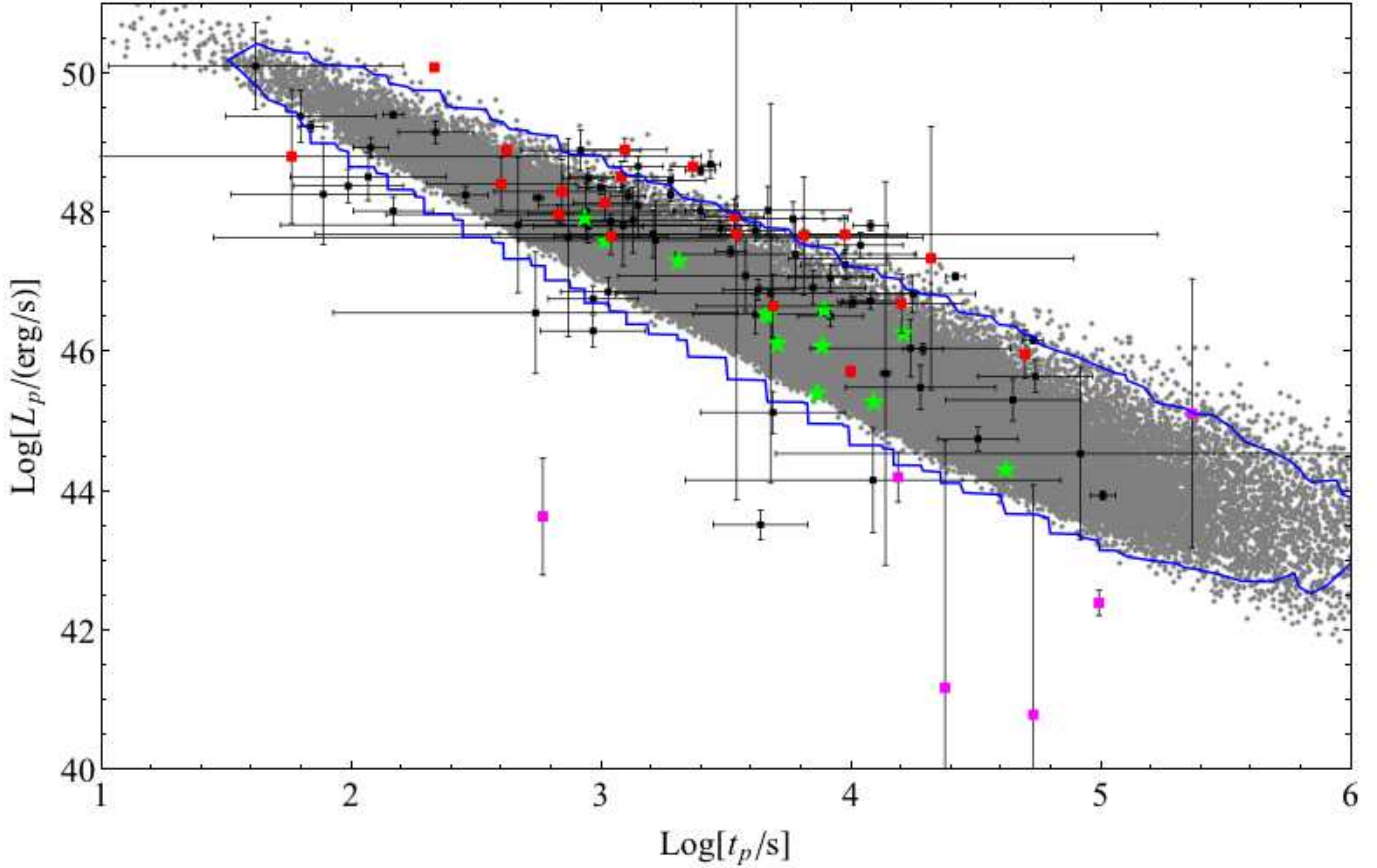}
\caption{Luminosity at the end time versus end time of the plateau phase. The black squares are the sample analysed by \citet{2010ApJ...722L.215D}. The red squares are Type Ia light curves of our sample and the pink squares are the luminosity of the last observation in Type Ib light curves in our sample. The grey dots are the $100000$ simulations of the luminosity at the spindown time and the spindown time assuming that the magnetic field and the NS period are normally distributed around the mean values we found for our sample. The blue line marks the region that includes $99\%$ of the simulations. The green stars are the values found for our best fit with \citet{2011A&A...526A.121D} model.}
\label{LTcorr}
\end{figure*}

The major advantage of this interpretation is that all the plateau properties are directly related to the central engine and, consequently, to the prompt emission. This explains the connection between the plateau energy and the prompt emission energy. The basic information can be derived from two main quantities: the magnetic field $B$ and the period $P$ of the pulsar. In particular, they are all that is needed to explain the anticorrelation between the plateau luminosity $L_p$ and the second break time $t_{b2}=t_p$ found by \citet{2008MNRAS.391L..79D,2010ApJ...722L.215D}. Starting from the distributions of $B$ and $P$ from our sample (see Appendix~\ref{app_magnetar}), we obtained the normalisation, the slope, and the scatter of the observed anticorrelation (Fig.~\ref{LTcorr}; for details see Appendix~\ref{app_magnetar}). We underline that these distributions are within the range of values expected for newly born, millisecond spinning magnetars \citep{1992ApJ...392L...9D}.

We plot in Fig.~\ref{LTcorr} the luminosity at the break time $L(t_{b1})$ of Type Ia light curves in the rest frame energy band corresponding to $\tilde{z}=2.29$, and we find that they follow the same luminosity-time anticorrelation and that they can be interpreted as the spin-down luminosity of a millisecond pulsar for the same distribution of $B$ and $P$ as for Type II. We add to Fig.~\ref{LTcorr} last Type Ib observation as a lower limit on the end of the injection phase and an upper limit on the luminosity. The possibility of having injection times up to $10^5$ s, as observed in Type Ib light curves, is allowed within reasonable values of the magnetic field and period (see Fig.~\ref{LTcorr}). However, the upper limit on luminosity found for some Type Ib is much lower than the expected one, unless we assume that the injection time is $\gtrsim 10^6$ s. We argue that a different beaming factor and/or efficiency in converting the spin-down power in X-rays may account for such Type Ib light curves.

The main constraints of the magnetar model is related to the energy budget. The maximum energy emitted in such a model is a few $10^{52}$ erg and limited by the maximum rotation energy attainable by a rotating neutron star \citep{1992Natur.357..472U}. The energy budget strongly depends on the uncertain estimate of the jet angle, however in a few cases the released energy may be high enough to challenge the model \citep{2010ApJ...711..641C}.

\subsubsection{The plateau phase in accretion models}

The plateau phase in the accretion model of \citet{2008MNRAS.388.1729K} is triggered if the progenitor star has a core-envelope structure. An analysis of Type II light curves with this model has been presented by \citet{2010MNRAS.401.1465C}. They inferred properties of the progenitor star based on the timescales and energetics, although they assume that the initial steep decay is not produced by the central engine. This approach is very effective in obtaining some indications about the progenitor star structure.

In Sect.~\ref{energy} we show that a quasi-linear correlation exists between shallow decay energy and the prompt emission total energy for the light curves with and without flares, extending the relation between the plateau energy and the prompt emission energy observed in $15-150$ keV previously found by \citet{2007ApJ...670..565L} \citep[see also][]{2010MNRAS.401.1465C}. It implies that the progenitor star has a self-similar structure, with a constant envelope-to-core mass ratio $\sim 0.02\div 0.03$. This is valid for Type II, as shown by \citet{2010MNRAS.401.1465C}, but also for Type Ia, where similar injection times (see Table~\ref{tab_median}) point to similar dimensions of the progenitor star. The slightly different distribution in the power-law indices of the two cases (see Table~\ref{tab_median}) can be ascribed to a possibly different typical accretion rate of the envelope (see e.g. \citealt{2008MNRAS.388.1729K}, their Fig.~7).

However, in Sect.~\ref{class} we show that the shallow decay in Type Ib is closely related to the one in Types Ia and II. This poses severe problems to this scenario. In fact, it is very difficult to obtain a shallow decay duration $\gtrsim 10^4$ s, since such an extended envelope is more probably ejected during the main burst \citep{2008MNRAS.388.1729K}.

\subsection{The normal decay}\label{FS}

\begin{figure}
\includegraphics[width=\hsize,clip]{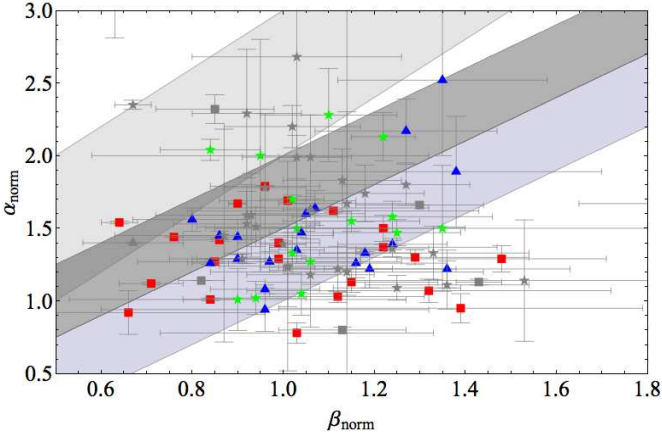}
\caption{Temporal decay index $\alpha_{norm}$ vs. spectral index $\beta_{norm}$ of Type 0 (red squares), of the second segment of Type Ia (blue triangles), and the third segment of Type II (green stars) light curves of the present sample. Grey symbols refer to the sample of \citet{2011MNRAS.410.1064M}. The shaded areas correspond to the expectations of the standard afterglow models in the slow cooling regime: the lower area only is for ISM model, the intermediate area plus the lower area are for the wind model \citep[see e.g. Table 2 in][]{2006ApJ...642..354Z}, the upper area is for the post jet-break \citep{1999ApJ...519L..17S}}
\label{abI}
\end{figure}

The forward shock emission is expected to give a significant contribution to the X-ray light curve after the fading of the prompt emission \citep[see e.g.][and references therein]{2006ApJ...642..354Z}. The shape of the light curve depends on the density profile of the ambient medium \citep{1998ApJ...497L..17S,2000ApJ...536..195C} and the temporal and spectral indices should satisfy specific closure relations \citep{2004IJMPA..19.2385Z}. Figure~\ref{abI} shows that $80\%$ of Type II light curves matches the expectations of the standard afterglow model in the slow cooling regime within $1 \sigma$, $90\%$ of the second segment of Type Ia, and $77\%$ of Type 0 light curves. This connection is favoured in Type Ia and II light curves by the absence of spectral variation during the transition from the previous injection phase (see Fig.~\ref{BETAtype}). If we include Type II GRBs of the sample analysed in \citet{2011MNRAS.410.1064M}, we find that $84\%$ are consistent with one of the above cases.

If the origin of the plateau phase is injection of energy from a millisecond pulsar into the forward shock \citep[][and references therein]{2011A&A...526A.121D}, after the plateau we expect that the light curves follow the standard afterglow scenario, as we find for the majority of cases. Alternatively, the accretion model predicts an asymptotic behaviour $\sim t^{-(1.3\div2.7)}$ after the end of the fall back of the envelope, which covers the range of variability of our $\alpha_{normal}$. However, this does not constrain the spectral index of the final decay to be related to the one of the previous accretion phase, as we found in Fig.~\ref{BETAtype}. On the contrary, since the late time decay should be produced by the same mechanism than the early steep decay, a similar spectral behaviour should be expected.

\subsection{The occurrence of flares}

The comparison of the properties of the present sample with the X-ray light curves of GRBs showing flaring activity in \citet{2011MNRAS.410.1064M} shows that flares occur more likely in Type II light curves ($73\%$ of the flare light curves are Type II, while only $23\%$ of the light curves without flares, as first noticed in \citealt{giantflares10}). We argue that the presence of the steep decay is related to suitable conditions for the occurrence of early-time flares. A connection between the flares and the steep decay has been found in \citet{2011MNRAS.410.1064M}: the average luminosity of flares decays in time as the average slope of the steep decay. In that work we also noted that light curves with multiple flares are associated with shallower decays. A hint of a steeper distribution of the power-law index for the light curves without flares is found in Fig.~\ref{alpha}, as if there is a transition from steep to shallow decays depending on the number of flares. However, the moderate difference is not conclusive.

The association of the steep decay with the flaring activity is an argument for a possible contribution of the central engine to the steep decay emission. In this case, the slope of the steep decay in the accretion model depends on the rotation rate of the progenitor core: the faster the rotation, the shallower the decay \citep[see Fig.~5 in][]{2008MNRAS.388.1729K}. A fast rotation of the infall material is needed to create the suitable conditions for instabilities, which are supposedly responsible for accretion models \citep{2006ApJ...636L..29P,2008MNRAS.388.1729K}. A change in the rotation rate of the core produces a change in the peak jet power and of the duration over which the luminosity is high, as shown in Fig.~5 of \citet{2008MNRAS.388.1729K}. However, we do not observe any significant difference in the $L_{pk}$ and $E_{iso}$ distributions of the GRBs with and without flares. 

In \citet{2011A&A...526A..27B} we showed that the decoupling of the evolution of the peak luminosity of late-time flares from the underlying continuum allows the detection of only the brightest flares. Comparing the average luminosity of the present sample with the one calculated for the continuum underlying the late-time flares analysed in \citet{2011A&A...526A..27B}, we observe that their temporal behaviour is similar, and it traces the flare detection threshold luminosity calculated in \citet{2011MNRAS.410.1064M}.

\section{Summary and conclusions}\label{conclusions}

We analysed $64$ long GRB X-ray light curves observed by XRT with redshift measurement that do not exhibit flaring activity. This allowed us to characterise the morphology and energetics of the sample in order to constrain the mechanism that produces the X-ray continuum. The light curves of the sample can be divided into four morphological Types, with similarities between different parts. When necessary, we compared our sample with the one considered in \citet{2011MNRAS.410.1064M} of $44$ GRBs that exhibit flaring activity. We found that

\begin{itemize}
\item a large fraction of the light curves are on average consistent with the high latitude emission from a structured jet, but all of them show a strong or moderate spectral evolution. The accretion model accounts for the observed range of temporal decays and the spectral evolution of the steep decay phases of Types Ib and II;
\item the injection phase observed in Type Ia and II can be interpreted as power emitted from spinning-down ultramagnetised neutron star that refreshes the forward shock. We applied such a model to our sample and found that it fits the data well enough;
\item this scenario accounts for the observed $L_p-t_p$ anticorrelation \citet{2008MNRAS.391L..79D,2010ApJ...722L.215D}: we reproduced the normalisation, slope, and scatter of the anticorrelation starting from plausible values of the parameters. The consistency of Type Ia shallow decay with the anticorrelation within $99\%$ confidence level suggests that they can be reproduced by this model with the same distribution of $B$ and $P$ as for Type II;
\item a second phase of accretion of a progenitor with a structure core envelope can originate the shallow phase in Types Ia and II. The correlation between the prompt emission total isotropic energy with the shallow decay X-ray energy implies a self-similar structure for the progenitor star, with $\sim 0.02-0.03$. However in this scenario, difficulties arise for the very long duration ($t\gtrsim 10^4$ s) of Type Ib shallow decay;
\item $70\%$ of the light curves with flares are Type II, while only $23\%$ of the non-flare sample are, confirming our previous finding \citep{giantflares10}. This suggests that Type II light curves are related to the suitable conditions for the occurrence of flares. The accretion model has the merit of proposing a unique mechanism to explain the steep decay and the flaring activity;
\end{itemize}

The magnetar model is able to explain the different morphologies of the X-ray continuum and the properties of the shallow decay. The main limit in the accretion model is the difficulty of achieving very long timescales as observed in Type Ib shallow decay. The behaviour of the steep decay and the suggested connection with the flaring activity \citep[see also][]{2009MNRAS.395..955B,giantflares10,2011MNRAS.410.1064M} indicates that the steep decay may not be viewed as simply the tail of the prompt emission.

\section*{Acknowledgments}
We thank the referee for her/his very useful comments and suggestions. This work is supported by ASI grant SWIFT I/011/07/0, by  the
Ministry of University and Research of Italy (PRIN MIUR 2007TNYZXL), by
MAE, and by the University of Milano Bicocca (Italy).

\appendix

\section{The best-fitting models}\label{function}

\textbf{Type 0} light curves correspond to a simple power-law model:
\begin{equation}
L(t)=N\,t^{-\alpha_1}\,.
\end{equation}

\textbf{Type I} light curves correspond to a smoothly-joined broken power-law model:
\begin{equation}
L(t)=N\left( \left( \frac{t}{t_{b1}}\right)^{\frac{\alpha_1}{d_1}}+ \left(\frac{t}{t_{b1}}\right)^{\frac{\alpha_2}{d_1}} \right)^{-d_1}.
\end{equation}

\textbf{Type II} light curves correspond to a smoothly-joined double-broken power-law model:
\begin{equation}
L(t)=N\left( \left( \frac{t}{t_{b1}}\right)^{\frac{\alpha_1}{d_1}}+ \left(\frac{t}{t_{b1}}\right)^{\frac{\alpha_2}{d_1}} \right)^{-d_1} \left( 1+ \left(\frac{t}{t_{b2}}\right)^{\frac{\alpha_3'}{d_2}} \right)^{-d_2}.
\label{DBPL}
\end{equation}
The $\alpha_3$ reported throughout the paper is not the best-fitting parameter $\alpha_3'$ that appears in Eq.~\ref{DBPL} but its asymptotic value $L(t\gg t_{b2})\propto t^{-\alpha_3}$.

\section{Energy injection from a spinning down neutron star}\label{app_magnetar}

The injection of energy into the forward shock from a millisecond spinning, ultramagnetised neutron star (NS) can be written as \citep{2011A&A...526A.121D}
\begin{equation}
\frac{dE}{dT}=L_{sd}(T)-k' \frac{E(T)}{T}=\frac{IK\omega^4_i}{(1+aT)^2}-k' \frac{E(T)}{T}
\end{equation}
where $I$ is the NS moment of inertia, $K=B^2R^6/(6Ic^3)$ with $B$ the dipole magnetic field at the NS pole, $R$ the NS radius, $c$ the speed of light, $a=1/t_2=2K\omega^2_i$ is the inverse of the spindown timescale, $k'$ is the radiative efficiency, and $E$ the initial blastwave energy. A solution of this equation is  \citep{2011A&A...526A.121D}
\begin{equation}
E(T)=\frac{IK\omega^4_i}{T^{k'}}\int^T_{T_\circ} \frac{T^{k'}}{(1+aT)^2}+E_\circ \left( \frac{T_\circ}{T} \right)^{k'}
\end{equation}
where $T_\circ$ is any time chosen as initial condition. The solution of the above integral can be expressed in terms of the real valued hypergeometric function $_{2}F_1(a,b,c;(1+aT)^{-1})$. The total bolometric luminosity of the blastwave is, then
\begin{equation}
 L(T)=k'E(T)/T\, .
\label{lum}
\end{equation}

\begin{figure}
\includegraphics[width=\hsize,clip]{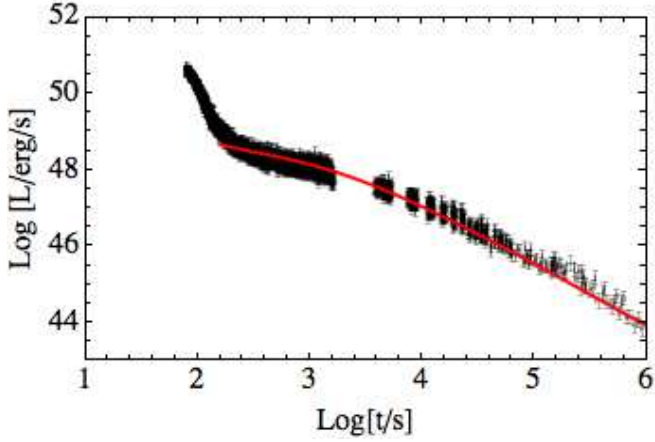}
\caption{GRB090618 light curve with the model in Eq.~\ref{lum} (red line) for the best-fit values reported in Table~\ref{tab_magnetar}.}
\label{fit_magnetar}
\end{figure}

We select $12$ Type II light curves from our sample that show a well-sampled, prominent plateau (see Table~\ref{tab_magnetar}), and we compute the $0.3-100$ keV energy band luminosity. We fit this sub sample with Eq.~\ref{lum}. We use as free parameters $B$, the NS period $P=1/\omega_i$, $E_\circ$, and $k'$. We adopt as starting time of the plateau phase $T_\circ$ the (rest-frame) time in which the spectral index $\beta$ becomes approximately constant. Figure~\ref{fit_magnetar} shows the result of the fit in one case, while the best-fit parameters for the entire sample can be found in Table~\ref{tab_magnetar}. We find that both the magnetic field and the NS period ($\bar{B}=3.4\times 10^{15}$ G; $\bar{P}=4.2$ ms) are within the range of values expected for newly born, millisecond spinning magnetars \citep{1992ApJ...392L...9D}, and consistent with the results obtained by \citet{2011A&A...526A.121D} and \citet{2010MNRAS.402..705L}. 

Assuming that the $B$ and $P$ are normally distributed around the mean values, we found for our sample $\bar{B}=3.4\times 10^{15}$ G and $Log[\bar{P}/ms]=0.62$, with standard deviations $\sigma_B=1.5\times 10^{15}$ G and $\sigma_{Log[P/ms]}=0.31$. We calculated the spindown timescale $t_2\propto B^2/P^2$ and the corresponding luminosity $L(t_2)$ according to Eq.~\ref{lum} ($T_\circ=10$ s, $k=\bar{k}$ and $E_\circ=\bar{E_\circ}$). Using an average spectral index $\tilde{\beta}=1$, we normalised the luminosity in the rest frame $0.3-10$ keV energy band corresponding to the average redshift of the \citet{2010ApJ...722L.215D} sample, $\tilde{z}=2.29$. The results are portrayed in Fig.~\ref{LTcorr}.

\begin{table}
 \begin{minipage}{85mm}
  \caption{Table of the Type II GRBs fitted with Eq.~\ref{lum}. $z$ is the redshift, $T_\circ$ the initial time, $B$ the best-fit values for magnetic field, $P$ for NS period, $k'$ for radiative efficiency and $E_\circ$ for blastwave energy. In the lower panel the same values are listed for a set of GRBs with flares (see also the fit of GRB060729 and GRB061121 in \citet{2011A&A...526A.121D}).}
  \label{tab_magnetar}
    \resizebox{\textwidth}{!}{
  \begin{tabular}{lllllll}
  GRB	&	z		&		$T_\circ$	(s) &		$B$ ($10^{15}$ G)	&		$P$ (ms)			&			$k'$				&		$E_\circ$ ($10^{50}$ erg)	\\
  \hline
051016B  	& $	0.9364	$ & $	600.		$ & $	7.0\pm 1.3		$ & $	11.0\pm 1.0		$ & $		0.80\pm 0.36		$ & $	    	  0.07\pm 1.3$	\\
051109A	& $	2.346	$ & $	100.		$ & $	2.8\pm 0.1		$ & $	1.5\pm0.1			$ & $		0.42\pm 0.06		$ & $	  	  10.0\pm 1.3$	\\
051221	& $	0.5465	$ & $	100.		$ & $	4.0\pm 0.0		$ & $   	15.0\pm 0.9		$ & $		0.52\pm 0.12		$ & $		  0.3\pm 0.2$	\\
060502A	& $	1.51		$ & $	400.		$ & $	2.2\pm 0.5		$ & $	2.7\pm 0.4		$ & $		0.25\pm 0.13		$ & $		  3.0\pm 0.8$	\\
060605	& $	3.78		$ & $	300.		$ & $	4.0\pm 1.8		$ & $	2.14\pm 0.2		$ & $		0.99\pm 0.56		$ & $	  	3.0\pm 1.5$	\\
070306	& $	1.4959	$ & $	1000.	$ & $	1.7\pm 8.9		$ & $	2.8\pm 0.2		$ & $		0.99\pm 0.33		$ & $		  1.5\pm 0.5$	\\
080707	& $	1.23		$ & $	100.		$ & $	5.0\pm 0.8		$ & $	10.2\pm 0.1		$ & $		0.99\pm 0.60		$ & $             		0.15\pm 0.36$	\\
090529	& $	2.625	$ & $	1000.	$ & $	1.2\pm 0.6		$ & $	2.8\pm 0.1		$ & $		0.89\pm 0.17		$ & $		  2.0\pm 0.8$	\\
090618	& $	0.54		$ & $	160.		$ & $	3.0\pm 0.0		$ & $	2.5\pm 0.0		$ & $		0.81\pm 0.01		$ & $		12.0\pm 5.0$	\\
090927	& $	1.37		$ & $	1000.	$ & $	3.2\pm 0.3		$ & $	5.1\pm 0.1		$ & $		0.99\pm 0.12		$ & $	          0.8\pm 1.4$	\\
100425A	& $	1.755	$ & $	700.		$ & $	4.0\pm 0.7		$ & $	5.2\pm 0.3		$ & $		0.51\pm 0.14		$ & $		  0.9\pm 0.8$	\\
100621	& $	0.542	$ & $	250.		$ & $	3.0\pm 0.0		$ & $	3.8\pm 0.0		$ & $		0.70\pm 0.02		$ & $		4.0\pm 0.3$	\\
\hline
060526  	& $	3.21	$ & $	100.		$ & $	3.2\pm 1.3		$ & $	5.5\pm 1.0		$ & $		0.70\pm 0.06		$ & $	    	  1.0\pm 0.6$	\\
060714  	& $	2.71	$ & $	60.		$ & $	5.0\pm 2.0		$ & $	2.8\pm 0.2		$ & $		0.43\pm 0.05		$ & $	10.0\pm 6.0    	  $	\\
080310  	& $	2.43	$ & $	400.		$ & $	6.3\pm 0.8		$ & $	5.8\pm 1.0		$ & $		0.80\pm 0.06		$ & $	    	  4.0\pm 1.2$	\\
091029  	& $	3.79	$ & $	100.	$ & $	1.99\pm 0.6		$ & $	1.6\pm 0.3		$ & $		0.32\pm 0.01		$ & $	1.6\pm 0.4    	  $	\\
\end{tabular}	} 	 
\end{minipage}														
\end{table}


\begin{thebibliography}{}

\bibitem[\protect\citeauthoryear{{Amati}, {Guidorzi}, {Frontera}, {Della
  Valle}, {Finelli}, {Landi} \& {Montanari}}{{Amati}
  et~al.}{2008}]{2008MNRAS.391..577A}
{Amati} L.,  {Guidorzi} C.,  {Frontera} F.,  {Della Valle} M.,  {Finelli} F.,
  {Landi} R.,    {Montanari} E.,  2008, MNRAS, 391, 577

\bibitem[\protect\citeauthoryear{{Barniol Duran} \& {Kumar}}{{Barniol Duran} \&
  {Kumar}}{2009}]{2009MNRAS.395..955B}
{Barniol Duran} R.,  {Kumar} P.,  2009, MNRAS, 395, 955

\bibitem[\protect\citeauthoryear{{Bernardini}, {Margutti}, {Chincarini},
  {Guidorzi} \& {Mao}}{{Bernardini} et~al.}{2011}]{2011A&A...526A..27B}
{Bernardini} M.~G.,  {Margutti} R.,  {Chincarini} G.,  {Guidorzi} C.,    {Mao}
  J.,  2011, A\&A, 526, A27

\bibitem[{{Bloom} {et~al.}(2001){Bloom}, {Frail}, \&
  {Sari}}]{2001AJ....121.2879B}
{Bloom}, J.~S., {Frail}, D.~A., \& {Sari}, R. 2001, AJ, 121, 2879

\bibitem[\protect\citeauthoryear{{Burrows}, {Hill}, {Nousek}, {Kennea},
  {Wells}, {Osborne}, {Abbey}, {Beardmore}, {Mukerjee}, {Short}, {Chincarini},
  {Campana}, {Citterio}, {Moretti}, {Pagani}, {Tagliaferri}, {Giommi},
  {Capalbi}, {Tamburelli}, {Angelini}}{{Burrows} et~al.}{2005}]{2005SSRv..120..165B}
{Burrows} D.~N. et al.,  2005, Space Science Reviews, 120, 165

\bibitem[\protect\citeauthoryear{{Butler}}{{Butler}}{2007}]{2007ApJ...656.1001%
B}
{Butler} N.~R.,  2007, ApJ, 656, 1001

\bibitem[\protect\citeauthoryear{{Butler} \& {Kocevski}}{{Butler} \&
  {Kocevski}}{2007}]{2007ApJ...663..407B}
{Butler} N.~R.,  {Kocevski} D.,  2007, ApJ, 663, 407

\bibitem[\protect\citeauthoryear{{Cenko}, {Frail}, {Harrison}, {Kulkarni},
  {Nakar}, {Chandra}, {Butler}, {Fox}, {Gal-Yam}, {Kasliwal}, {Kelemen},
  {Moon}, {Ofek}, {Price}, {Rau}, {Soderberg}, {Teplitz}, {Werner}, {Bock},
  {Bloom}, {Starr}, {Filippenko}}{{Cenko} et~al.}{2010}]{2010ApJ...711..641C}
{Cenko} S.~B., et al.  2010, ApJ,
  711, 641

\bibitem[\protect\citeauthoryear{{Chevalier} \& {Li}}{{Chevalier} \&
  {Li}}{2000}]{2000ApJ...536..195C}
{Chevalier} R.~A.,  {Li} Z.,  2000, ApJ, 536, 195

\bibitem[\protect\citeauthoryear{{Chincarini}, {Mao}, {Margutti}, {Bernardini},
  {Guidorzi}, {Pasotti}, {Giannios}, {Della Valle}, {Moretti}, {Romano},
  {D'Avanzo}, {Cusumano} \& {Giommi}}{{Chincarini} et~al.}{2010}]{chinca10}
{Chincarini} G.,  {Mao} J.,  {Margutti} R.,  {Bernardini} M.~G.,  {Guidorzi}
  C.,  {Pasotti} F.,  {Giannios} D.,  {Della Valle} M.,  {Moretti} A.,
  {Romano} P.,  {D'Avanzo} P.,  {Cusumano} G.,    {Giommi} P.,  2010, MNRAS,
  406, 2113


\bibitem[\protect\citeauthoryear{{Chincarini}, {Moretti}, {Romano}, {Falcone},
  {Morris}, {Racusin}, {Campana}, {Covino}, {Guidorzi}, {Tagliaferri},
  {Burrows}, {Pagani}, {Stroh}, {Grupe}, {Capalbi}, {Cusumano}, {Gehrels},
  {Giommi}, {La Parola} \& {Mangano}}{{Chincarini} et~al.}{2007}]{2007ApJ...671.1903C}
{Chincarini} G. et al.,
   2007, ApJ, 671, 1903

\bibitem[\protect\citeauthoryear{{Corsi} \& {M{\'e}sz{\'a}ros}}{{Corsi} \&
  {M{\'e}sz{\'a}ros}}{2009}]{2009ApJ...702.1171C}
{Corsi} A.,  {M{\'e}sz{\'a}ros} P.,  2009, ApJ, 702, 1171

\bibitem[\protect\citeauthoryear{{Cui}, {Liang}, {Lv}, {Zhang} \& {Xu}}{{Cui}
  et~al.}{2010}]{2010MNRAS.401.1465C}
{Cui} X.,  {Liang} E.,  {Lv} H.,  {Zhang} B.,    {Xu} R.,  2010, MNRAS, 401,
  1465

\bibitem[\protect\citeauthoryear{{Curran}, {Starling}, {O'Brien}, {Godet}, {van
  der Horst} \& {Wijers}}{{Curran} et~al.}{2008}]{2008A&A...487..533C}
{Curran} P.~A.,  {Starling} R.~L.~C.,  {O'Brien} P.~T.,  {Godet} O.,  {van der
  Horst} A.~J.,    {Wijers} R.~A.~M.~J.,  2008, A\&A, 487, 533

\bibitem[\protect\citeauthoryear{{Dai} \& {Lu}}{{Dai} \&
  {Lu}}{1998}]{1998A&A...333L..87D}
{Dai} Z.~G.,  {Lu} T.,  1998, A\&A, 333, L87

\bibitem[\protect\citeauthoryear{{Dainotti}, {Cardone} \&
  {Capozziello}}{{Dainotti} et~al.}{2008}]{2008MNRAS.391L..79D}
{Dainotti} M.~G.,  {Cardone} V.~F.,    {Capozziello} S.,  2008, MNRAS, 391, L79

\bibitem[\protect\citeauthoryear{{Dainotti}, {Willingale}, {Capozziello},
  {Fabrizio Cardone} \& {Ostrowski}}{{Dainotti}
  et~al.}{2010}]{2010ApJ...722L.215D}
{Dainotti} M.~G.,  {Willingale} R.,  {Capozziello} S.,  {Fabrizio Cardone} V.,
    {Ostrowski} M.,  2010, ApJ, 722, L215

\bibitem[\protect\citeauthoryear{{Dall'Osso}, {Stratta}, {Guetta}, {Covino},
  {de Cesare} \& {Stella}}{{Dall'Osso} et~al.}{2011}]{2011A&A...526A.121D}
{Dall'Osso} S.,  {Stratta} G.,  {Guetta} D.,  {Covino} S.,  {de Cesare} G.,
  {Stella} L.,  2011, A\&A, 526, A121

\bibitem[\protect\citeauthoryear{{Duncan} \& {Thompson}}{{Duncan} \&
  {Thompson}}{1992}]{1992ApJ...392L...9D}
{Duncan} R.~C.,  {Thompson} C.,  1992, ApJ, 392, L9

\bibitem[\protect\citeauthoryear{{Evans}, {Beardmore}, {Page}, {Osborne},
  {O'Brien}, {Willingale}, {Starling}, {Burrows}, {Godet}, {Vetere}, {Racusin},
  {Goad}, {Wiersema}, {Angelini}, {Capalbi}, {Chincarini}, {Gehrels}, {Kennea},
  {Margutti}, {Morris}}{{Evans} et~al.}{2009}]{2009MNRAS.397.1177E}
{Evans} P.~A. et al.,  2009, MNRAS, 397, 1177

\bibitem[\protect\citeauthoryear{{Fan} \& {Piran}}{{Fan} \&
  {Piran}}{2006}]{2006MNRAS.369..197F}
{Fan} Y.,  {Piran} T.,  2006, MNRAS, 369, 197

\bibitem[\protect\citeauthoryear{{Gehrels}, {Barthelmy}, {Burrows}, {Cannizzo},
  {Chincarini}, {Fenimore}, {Kouveliotou}, {O'Brien}, {Palmer}, {Racusin},
  {Roming}, {Sakamoto}, {Tueller}, {Wijers} \& {Zhang}}{{Gehrels}
  et~al.}{2008}]{2008ApJ...689.1161G}
{Gehrels} N.,  {Barthelmy} S.~D.,  {Burrows} D.~N.,  {Cannizzo} J.~K.,
  {Chincarini} G.,  {Fenimore} E.,  {Kouveliotou} C.,  {O'Brien} P.,  {Palmer}
  D.~M.,  {Racusin} J.,  {Roming} P.~W.~A.,  {Sakamoto} T.,  {Tueller} J.,
  {Wijers} R.~A.~M.~J.,    {Zhang} B.,  2008, ApJ, 689, 1161

\bibitem[\protect\citeauthoryear{{Gehrels}, {Chincarini}, {Giommi}, {Mason},
  {Nousek}, {Wells}, {White}, {Barthelmy}, {Burrows}, {Cominsky}, {Hurley},
  {Marshall}, {M{\'e}sz{\'a}ros}, {Roming}, {Angelini}, {Barbier}, {Belloni},
  {Campana}, {Caraveo}, {Chester}}{{Gehrels}
  et~al.}{2004}]{2004ApJ...611.1005G}
{Gehrels} N. et al.,  2004, ApJ, 611, 1005

\bibitem[\protect\citeauthoryear{{Krimm}, {Yamaoka}, {Sugita}, {Ohno},
  {Sakamoto}, {Barthelmy}, {Gehrels}, {Hara}, {Norris}, {Ohmori}, {Onda},
  {Sato}, {Tanaka}, {Tashiro} \& {Yamauchi}}{{Krimm}
  et~al.}{2009}]{2009ApJ...704.1405K}
{Krimm} H.~A.,  {Yamaoka} K.,  {Sugita} S.,  {Ohno} M.,  {Sakamoto} T.,
  {Barthelmy} S.~D.,  {Gehrels} N.,  {Hara} R.,  {Norris} J.~P.,  {Ohmori} N.,
  {Onda} K.,  {Sato} G.,  {Tanaka} H.,  {Tashiro} M.,    {Yamauchi} M.,  2009,
  ApJ, 704, 1405

\bibitem[\protect\citeauthoryear{{Kumar}, {Narayan} \& {Johnson}}{{Kumar}
  et~al.}{2008}]{2008MNRAS.388.1729K}
{Kumar} P.,  {Narayan} R.,    {Johnson} J.~L.,  2008, MNRAS, 388, 1729

\bibitem[\protect\citeauthoryear{{Kumar} \& {Panaitescu}}{{Kumar} \&
  {Panaitescu}}{2000}]{2000ApJ...541L..51K}
{Kumar} P.,  {Panaitescu} A.,  2000, ApJ, 541, L51

\bibitem[\protect\citeauthoryear{{Liang}, {L{\"u}}, {Hou}, {Zhang} \&
  {Zhang}}{{Liang} et~al.}{2009}]{2009ApJ...707..328L}
{Liang} E.,  {L{\"u}} H.,  {Hou} S.,  {Zhang} B.,    {Zhang} B.,  2009, ApJ,
  707, 328

\bibitem[\protect\citeauthoryear{{Liang}, {Racusin}, {Zhang}, {Zhang} \&
  {Burrows}}{{Liang} et~al.}{2008}]{2008ApJ...675..528L}
{Liang} E.,  {Racusin} J.~L.,  {Zhang} B.,  {Zhang} B.,    {Burrows} D.~N.,
  2008, ApJ, 675, 528

\bibitem[\protect\citeauthoryear{{Liang}, {Zhang} \& {Zhang}}{{Liang}
  et~al.}{2007}]{2007ApJ...670..565L}
{Liang} E.,  {Zhang} B.,    {Zhang} B.,  2007, ApJ, 670, 565

\bibitem[\protect\citeauthoryear{{Lindner}, {Milosavljevi{\'c}}, {Couch} \&
  {Kumar}}{{Lindner} et~al.}{2010}]{2010ApJ...713..800L}
{Lindner} C.~C.,  {Milosavljevi{\'c}} M.,  {Couch} S.~M.,    {Kumar} P.,  2010,
  ApJ, 713, 800

\bibitem[\protect\citeauthoryear{{Lyons}, {O'Brien}, {Zhang}, {Willingale},
  {Troja} \& {Starling}}{{Lyons} et~al.}{2010}]{2010MNRAS.402..705L}
{Lyons} N.,  {O'Brien} P.~T.,  {Zhang} B.,  {Willingale} R.,  {Troja} E.,
  {Starling} R.~L.~C.,  2010, MNRAS, 402, 705

\bibitem[\protect\citeauthoryear{{Lyutikov}}{{Lyutikov}}{2006}]{2006MNRAS.369L...5L}
{Lyutikov} M.,  2006, MNRAS, 369, L5

\bibitem[\protect\citeauthoryear{{Lyutikov}}{{Lyutikov}}{2009}]{2009arXiv0911.0349L}
{Lyutikov} M.,  2009, ArXiv 0911.0349

\bibitem[\protect\citeauthoryear{{Lyutikov} \& {Blandford}}{{Lyutikov} \&
  {Blandford}}{2003}]{2003astro.ph.12347L}
{Lyutikov} M.,  {Blandford} R.,  2003, astro-ph/0312347

\bibitem[\protect\citeauthoryear{{Margutti}, {Bernardini}, {Barniol  Duran}, {Guidorzi},
  {Shen} \& {Chincarini}}{{Margutti} et~al.}{2011}]{2011MNRAS.410.1064M}
{Margutti} R.,  {Bernardini} G., {Barniol  Duran} R.,  {Guidorzi} C.,  {Shen} R.~F.,
     {Chincarini} G.,  2011, MNRAS, 410, 1064

\bibitem[\protect\citeauthoryear{{Margutti}, {Genet}, {Granot}, {Barniol
  Duran}, {Guidorzi}, {Chincarini}, {Mao}, {Schady}, {Sakamoto}, {Miller},
  {Olofsson}, {Bloom}, {Evans}, {Fynbo}, {Malesani}, {Moretti}, {Pasotti},
  {Starr}, {Burrows}, {Barthelmy}}{{Margutti} et~al.}{2010}]{2010MNRAS.402...46M}
{Margutti} R. et al.,  2010, MNRAS, 402, 46

\bibitem[\protect\citeauthoryear{{Margutti}, {Guidorzi}, {Chincarini},
  {Bernardini}, {Genet}, {Mao} \& {Pasotti}}{{Margutti}
  et~al.}{2010}]{giantflares10}
{Margutti} R.,  {Guidorzi} C.,  {Chincarini} G.,  {Bernardini} M.~G.,  {Genet}
  F.,  {Mao} J.,    {Pasotti} F.,  2010, MNRAS, 406, 2149

\bibitem[\protect\citeauthoryear{{Metzger}, {Giannios}, {Thompson},
  {Bucciantini} \& {Quataert}}{{Metzger} et~al.}{2010}]{2010arXiv1012.0001M}
{Metzger} B.~D.,  {Giannios} D.,  {Thompson} T.~A.,  {Bucciantini} N.,
  {Quataert} E.,  2011, MNRAS, 413, 2031

\bibitem[\protect\citeauthoryear{{Nava}, {Ghirlanda}, {Ghisellini} \&
  {Celotti}}{{Nava} et~al.}{2010}]{2010arXiv1012.3968N}
{Nava} L.,  {Ghirlanda} G.,  {Ghisellini} G., {Celotti} A.,  2011, A\&A, 530, A21

\bibitem[\protect\citeauthoryear{{Nousek}, {Kouveliotou}, {Grupe}, {Page},
  {Granot}, {Ramirez-Ruiz}, {Patel}, {Burrows}, {Mangano}, {Barthelmy},
  {Beardmore}, {Campana}, {Capalbi}, {Chincarini}, {Cusumano}, {Falcone},
  {Gehrels}, {Giommi}, {Goad}, {Godet}}{{Nousek} et~al.}{2006}]{2006ApJ...642..389N}
{Nousek} J.~A. et al.,  2006, ApJ, 642, 389

\bibitem[\protect\citeauthoryear{{O'Brien}, {Willingale}, {Osborne}, {Goad},
  {Page}, {Vaughan}, {Rol}, {Beardmore}, {Godet}, {Hurkett}, {Wells}, {Zhang},
  {Kobayashi}, {Burrows}, {Nousek}, {Kennea}, {Falcone}, {Grupe}, {Gehrels},
  {Barthelmy}, {Cannizzo}}{{O'Brien} et~al.}{2006}]{2006ApJ...647.1213O}
{O'Brien} P.~T. et al.,  2006, ApJ, 647, 1213

\bibitem[\protect\citeauthoryear{{Panaitescu} \& {Vestrand}}{{Panaitescu} \&
  {Vestrand}}{2011}]{2011MNRAS.414.3537P}
{Panaitescu} A., {Vestrand} W.~T.,  2011, MNRAS, 414, 3537

\bibitem[\protect\citeauthoryear{{Perna}, {Armitage} \& {Zhang}}{{Perna}
  et~al.}{2006}]{2006ApJ...636L..29P}
{Perna} R.,  {Armitage} P.~J.,    {Zhang} B.,  2006, ApJ, 636, L29

\bibitem[\protect\citeauthoryear{{Racusin}, {Liang}, {Burrows}, {Falcone},
  {Sakamoto}, {Zhang}, {Zhang}, {Evans} \& {Osborne}}{{Racusin}
  et~al.}{2009}]{2009ApJ...698...43R}
{Racusin} J.~L.,  {Liang} E.~W.,  {Burrows} D.~N.,  {Falcone} A.,  {Sakamoto}
  T.,  {Zhang} B.~B.,  {Zhang} B.,  {Evans} P.,    {Osborne} J.,  2009, ApJ,
  698, 43

\bibitem[\protect\citeauthoryear{{Romano}, {Campana}, {Chincarini}, {Cummings},
  {Cusumano}, {Holland}, {Mangano}, {Mineo}, {Page}, {Pal'Shin}, {Rol},
  {Sakamoto}, {Zhang}, {Aptekar}, {Barbier}, {Barthelmy}, {Beardmore}, {Boyd},
  {Burrows}, {Capalbi}}{{Romano} et~al.}{2006}]{2006A&A...456..917R}
{Romano} P. et al.,  2006,
  A\&A, 456, 917

\bibitem[\protect\citeauthoryear{{Sari}, {Piran} \& {Halpern}}{{Sari}
  et~al.}{1999}]{1999ApJ...519L..17S}
{Sari} R.,  {Piran} T.,    {Halpern} J.~P.,  1999, ApJ, 519, L17

\bibitem[\protect\citeauthoryear{{Sari}, {Piran} \& {Narayan}}{{Sari}
  et~al.}{1998}]{1998ApJ...497L..17S}
{Sari} R.,  {Piran} T.,    {Narayan} R.,  1998, ApJ, 497, L17

\bibitem[\protect\citeauthoryear{{Thompson}}{{Thompson}}{2006}]{2006ApJ...651.%
.333T}
{Thompson} C.,  2006, ApJ, 651, 333

\bibitem[\protect\citeauthoryear{{Usov}}{{Usov}}{1992}]{1992Natur.357..472U}
{Usov} V.~V.,  1992, nat, 357, 472

\bibitem[\protect\citeauthoryear{{Vaughan}, {Goad}, {Beardmore}, {O'Brien},
  {Osborne}, {Page}, {Barthelmy}, {Burrows}, {Campana}, {Cannizzo}, {Capalbi},
  {Chincarini}, {Cummings}, {Cusumano}, {Giommi}, {Godet}, {Hill}, {Kobayashi},
  {Kumar}, {La Parola}}{{Vaughan} et~al.}{2006}]{2006ApJ...638..920V}
{Vaughan} S. et al.,  2006, ApJ, 638, 920

\bibitem[\protect\citeauthoryear{{Willingale}, {O'Brien}, {Osborne}, {Godet},
  {Page}, {Goad}, {Burrows}, {Zhang}, {Rol}, {Gehrels} \&
  {Chincarini}}{{Willingale} et~al.}{2007}]{2007ApJ...662.1093W}
{Willingale} R.,  {O'Brien} P.~T.,  {Osborne} J.~P.,  {Godet} O.,  {Page}
  K.~L.,  {Goad} M.~R.,  {Burrows} D.~N.,  {Zhang} B.,  {Rol} E.,  {Gehrels}
  N.,    {Chincarini} G.,  2007, ApJ, 662, 1093

\bibitem[{{Zaninoni} {et~al.}(2011){Zaninoni}, {Margutti}, {Grazia Bernardini},
  \& {Chincarini}}]{2011arXiv1107.2870Z}
{Zaninoni}, E., {Margutti}, R., {Bernardini}, M.G., \& {Chincarini}, G.
  2011, ArXiv1107.2870

\bibitem[\protect\citeauthoryear{{Zhang}, {Fan}, {Dyks}, {Kobayashi},
  {M{\'e}sz{\'a}ros}, {Burrows}, {Nousek} \& {Gehrels}}{{Zhang}
  et~al.}{2006}]{2006ApJ...642..354Z}
{Zhang} B.,  {Fan} Y.~Z.,  {Dyks} J.,  {Kobayashi} S.,  {M{\'e}sz{\'a}ros} P.,
  {Burrows} D.~N.,  {Nousek} J.~A.,    {Gehrels} N.,  2006, ApJ, 642, 354

\bibitem[\protect\citeauthoryear{{Zhang}, {Liang} \& {Zhang}}{{Zhang}
  et~al.}{2007}]{2007ApJ...666.1002Z}
{Zhang} B.,  {Liang} E.,    {Zhang} B.,  2007, ApJ, 666, 1002

\bibitem[\protect\citeauthoryear{{Zhang} \& {M{\'e}sz{\'a}ros}}{{Zhang} \&
  {M{\'e}sz{\'a}ros}}{2001}]{2001ApJ...552L..35Z}
{Zhang} B.,  {M{\'e}sz{\'a}ros} P.,  2001, ApJ, 552, L35

\bibitem[\protect\citeauthoryear{{Zhang} \& {M{\'e}sz{\'a}ros}}{{Zhang} \&
  {M{\'e}sz{\'a}ros}}{2002}]{2002ApJ...571..876Z}
{Zhang} B.,  {M{\'e}sz{\'a}ros} P.,  2002, ApJ, 571, 876

\bibitem[\protect\citeauthoryear{{Zhang} \& {M{\'e}sz{\'a}ros}}{{Zhang} \&
  {M{\'e}sz{\'a}ros}}{2004}]{2004IJMPA..19.2385Z}
{Zhang} B.,  {M{\'e}sz{\'a}ros} P.,  2004, International Journal of Modern
  Physics A, 19, 2385

\end{thebibliography}
\end{document}